\documentclass[conference]{IEEEtran}
\IEEEoverridecommandlockouts
\usepackage[ruled,vlined,linesnumbered]{algorithm2e}
\usepackage{cite}
\usepackage{amsmath,amssymb,amsfonts}
\usepackage{algorithmic}
\usepackage{graphicx} 
\usepackage{textcomp} 
\usepackage{xcolor}
\usepackage{colortbl}
\usepackage{hyperref}
\usepackage{subcaption} 
\usepackage{booktabs} 
\usepackage{multirow} 
\usepackage{caption} 
   
\def\BibTeX{{\rm B\kern-.05em{\sc i\kern-.025em b}\kern-.08em
    T\kern-.1667em\lower.7ex\hbox{E}\kern-.125emX}}

\begin{document}

\title{CO-EVOLVE: Bidirectional Co-Evolution of Graph Structure and Semantics for Heterophilous Learning}

\author{\IEEEauthorblockN{Jinming Xing}
    \textit{North Carolina State University}\\
    Raleigh, NC, USA \\
    jxing6@ncsu.edu
    \and
    \IEEEauthorblockN{Muhammad Shahzad}
    \textit{North Carolina State University}\\
    Raleigh, NC, USA \\
    mshahzad@ncsu.edu
}

\maketitle

\sloppy
\begin{abstract}
    The integration of Large Language Models (LLMs) and Graph Neural Networks (GNNs) promises to unify semantic understanding with structural reasoning, yet existing methods typically rely on static, unidirectional pipelines. These approaches suffer from fundamental limitations: (1) Bidirectional Error Propagation, where semantic hallucinations in LLMs or structural noise in GNNs permanently poison the downstream modality without opportunity for recourse; (2) Semantic-Structural Dissonance, particularly in heterophilous settings where textual similarity contradicts topological reality; and (3) a Blind Leading the Blind phenomenon, where indiscriminate alignment forces models to mirror each other's mistakes regardless of uncertainty. To address these challenges, we propose CO-EVOLVE, a dual-view co-evolution framework that treats graph topology and semantic embeddings as dynamic, mutually reinforcing latent variables. By employing a Gauss-Seidel alternating optimization strategy, our framework establishes a cyclic feedback loop: the GNN injects structural context as Soft Prompts to guide the LLM, while the LLM constructs favorable Dynamic Semantic Graphs to rewire the GNN. We introduce three key innovations to stabilize this evolution: (1) a Hard-Structure Conflict-Aware Contrastive Loss that warps the semantic manifold to respect high-order topological boundaries; (2) an Adaptive Node Gating Mechanism that dynamically fuses static and learnable structures to recover missing links; and (3) an Uncertainty-Gated Consistency strategy that enables meta-cognitive alignment, ensuring models only learn from the confident view. Finally, an Entropy-Aware Adaptive Fusion integrates predictions during inference. Extensive experiments on public benchmarks demonstrate that CO-EVOLVE significantly outperforms state-of-the-art baselines, achieving average improvements of 9.07\% in Accuracy and 7.19\% in F1-score.
\end{abstract}

\section{Introduction}
Graph Neural Networks (GNNs) have established themselves as the de facto standard for modeling structural dependencies in relational data~\cite{kipf2017semi,ju2025survey,shehzad2026graph}, while Large Language Models (LLMs) have revolutionized the processing of semantic information~\cite{vaswani2017attention,brown2020language,liu2024deepseek,wang2025comprehensive}. Consequently, a rapidly growing body of research seeks to integrate these two powerful modalities, aiming to leverage the complementary strengths of structural reasoning and semantic understanding~\cite{ren2024survey,wang2025graph}. This intersection is particularly crucial for complex, knowledge-intensive domains, such as citation networks~\cite{chen2025beyond}, social media analysis~\cite{yang2025flag}, and molecular discovery~\cite{wang2025bridging}, where neither topology nor text alone captures the full underlying manifold. Recent advancements have explored this synergy from multiple angles. For instance, TAPE~\cite{he2024harnessing} utilizes LLMs to generate explanation-augmented node features, significantly enhancing GNN performance on text-attributed graphs by providing richer semantic context. In terms of unifying diverse graph tasks, OFA~\cite{liu2023one} proposes a general framework that describes structural information in natural language, enabling a single LLM to handle varied graph challenges across domains. Specific applications also illustrate the power of this integration: LLMRec~\cite{wei2024llmrec} employs text-generated profiles to filter noisy interactions and enrich node embeddings in recommender systems, while GAugLLM~\cite{fang2024gaugl} leverages LLMs for data augmentation in graph contrastive learning, addressing the limitations of traditional methods that overlook textual semantics. FLAG~\cite{yang2025flag} investigates financial fraud detection by integrating LLMs with GNNs through semantic neighbor sampling and discriminative text extraction. PromptGFM~\cite{zhu2025llm} proposes a unified framework that combines graph vocabulary learning with LLMs to support a wide range of downstream graph tasks.

\textbf{Limitations of Prior Art.} Despite the promise of this convergence, recent SOTA methods typically adopt a static unidirectional approach that fails to fully capture the dynamic interplay between structure and semantics. Early works like GIANT~\cite{chien2021node}, EdgeFormers~\cite{jin2023edgeformers}, and OFA~\cite{liu2023one} follow a \texttt{Text} $\rightarrow$ \texttt{LLM} $\rightarrow$ \texttt{GNN} pipeline, treating LLMs merely as fixed feature encoders. Conversely, recent GNN-as-Prefix approaches such as LLaGA~\cite{chen2024llaga}, GraphAdapter~\cite{huang2024can}, and InstructGLM~\cite{ye2024language} employ a \texttt{GNN} $\rightarrow$ \texttt{Projector} $\rightarrow$ \texttt{LLM} pipeline, injecting structural encodings into frozen LLMs. However, these static paradigms suffer from three fundamental limitations that hinder their performance, particularly on complex and heterophilous datasets: \textbf{First, Bidirectional Error Propagation:} Regardless of direction, these pipelines treat the output of the first stage as a fixed ground truth. In LLM-first models like GIANT~\cite{chien2021node}, TAPE~\cite{he2024harnessing}, and ZeroG~\cite{li2024zerog}, semantic hallucinations or irrelevant keywords permanently poison the downstream GNN. In GNN-first models like InstructGLM~\cite{ye2024language}, GraphGPT~\cite{tang2023graphgpt}, and LLaGA~\cite{chen2024llaga}, structural noise, such as missing edges or heterophilous connections, permanently biases the LLM's reasoning. Crucially, neither approach allows for \textit{mutual correction}, leading to unrecoverable error propagation where the downstream model cannot signal the upstream model to revise its representations. \textbf{Second, Semantic-Structural Dissonance:} Current co-training methods like FLAG~\cite{yang2025flag}, Patton~\cite{jin2023patton}, and PromptGFM~\cite{zhu2025llm} often assume Homophily, that semantic similarity and structural connectivity are perfectly aligned. They rely on global contrastive losses that collapse decision boundaries by treating all semantic neighbors as positive pairs. This fails in real-world scenarios characterized by (1) Heterophily, where nodes share keywords but belong to opposing structural communities, and (b) Sparsity, where semantically identical nodes lack explicit connections, causing GNNs to under-smooth representations. \textbf{Third, Blind Alignment:} Consistency regularization techniques \cite{zhang2021scr,yang2025flag} typically force the LLM and GNN probability distributions to match indiscriminately using KL divergence. This typically ignores model confidence, resulting in a Blind Leading the Blind scenario: when one modality is uncertain or confidently wrong, forcing the other to align with it simply propagates noise, degrading robustness in noisy environments.

\textbf{Problem Statement.} These limitations fundamentally stem from treating structure and semantics as static, independent inputs rather than dynamic, interacting latent variables. The core challenge is to break the Static Pipeline cycle to prevent bidirectional error propagation, while simultaneously resolving the Semantic-Structural Dissonance that arises when textual and topological signals conflict. Furthermore, effective integration requires moving beyond Blind Alignment to a meta-cognitive strategy where models only learn from each other when confident.

\textbf{Proposed Solution.} To address these challenges, we propose \textbf{CO-EVOLVE}, a Dual-View Co-Evolution Framework that unifies LLMs and GNNs into a mutually reinforcing cycle. Unlike static pipelines, we treat Graph Topology and Semantic Embeddings as latent variables that evolve together via Gauss-Seidel alternating optimization. Specifically, we dismantle the Static Pipeline by establishing a bidirectional feedback loop: the GNN injects topological context as Soft Prompts to guide the LLM, reducing hallucinations, while the LLM generates evolving embeddings to construct a Dynamic Semantic Graph that rewires the GNN, correcting structural noise. To resolve Semantic-Structural Dissonance on heterophilous graphs, we introduce a Hard-Structure Conflict-Aware Contrastive Loss that warps the semantic manifold to respect high-order topological boundaries, alongside an Adaptive Node Gating Mechanism that dynamically reconstructs reliable structure from evolving semantics. Finally, to mitigate Blind Alignment, we employ an Uncertainty-Gated Consistency strategy that acts as a meta-cognitive filter, ensuring models only align with the counter-view when it demonstrates high confidence. Extensive experiments on public datasets validate the effectiveness of CO-EVOLVE, with average improvements of 9.07\% in Accuracy and 7.19\% in F1-score.

\textbf{Contributions.} We summarize our primary contributions as follows:

\begin{itemize}
    \item   \textbf{Dual-View Co-Evolution with Bidirectional Feedback:} We introduce a cyclic interaction mechanism where the GNN injects topological context as Soft Prompts to guide the LLM, while the LLM simultaneously generates evolving embeddings to construct a Dynamic Semantic Graph for the GNN. This halts bidirectional error propagation.
    \item   \textbf{Adaptive Semantic-Guided Structure Learning:} We propose a Node-Adaptive Gating Mechanism that learns a per-node factor $\alpha_i$ to dynamically fuse the static and dynamic graphs. This resolves the Missing Link problem by reconstructing connections for isolated nodes while preserving the static graph where it is robust.
    \item   \textbf{Hard-Structure Conflict-Aware Loss:} To handle heterophily, we introduce a novel objective based on Global Graph Diffusion (PPR). By explicitly mining Hard Conflict Negatives, pairs that are semantically proximal but topologically distant, we force the semantic manifold to respect high-order structural boundaries.
    \item   \textbf{Uncertainty-Gated Consistency:} We develop a meta-cognitive alignment strategy weighted by the entropy of the teacher. This solves the Blind Alignment problem by ensuring the LLM only learns from the GNN when the GNN is confident, and vice-versa.
    \item   Extensive experiments on public datasets the superiority of CO-EVOLVE, with average improvements of 9.07\% in Accuracy and 7.19\% in F1-score.
\end{itemize}

\begin{figure*}[h!]
    \centering
    \includegraphics[width=\textwidth]{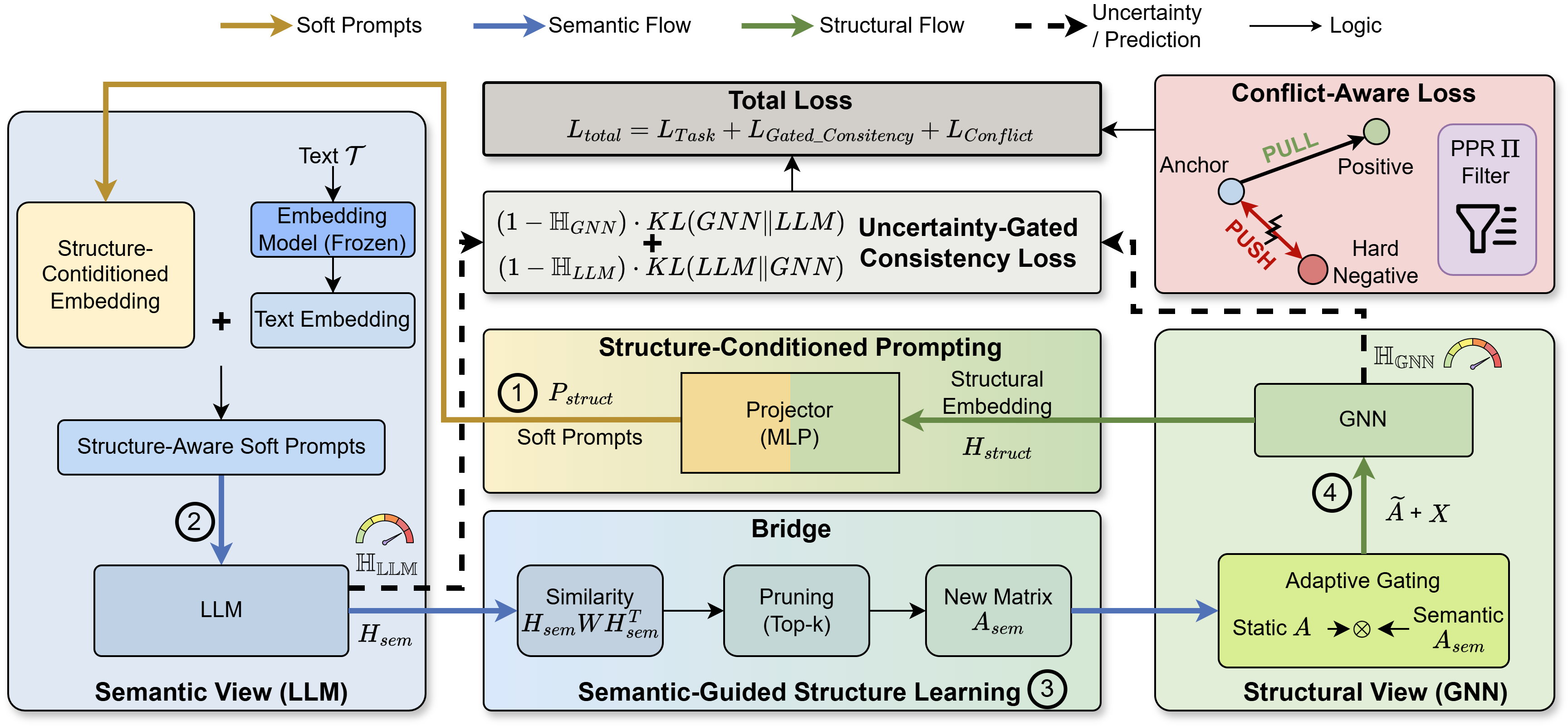}
    \caption{\textbf{The CO-EVOLVE Framework}. The training follows a sequential cycle: (1) The GNN encodes structural context into \textbf{Soft Prompts} to guide the LLM. (2) The LLM, conditioned on these prompts, generates semantic embeddings. (3) A \textbf{Semantic-Guided Structure Learning module} refines the graph topology based on the updated semantics, enhanced by the \textbf{Adaptive Gating}. (4) The GNN updates its representations using the refined graph. The training is supervised by an \textbf{Uncertainty-Gated Consistency Loss, a Hard-Structure Conflict Loss}, and a classification loss, ensuring robust alignment between modalities.}
    \label{fig:framework}
\end{figure*}
\section{Preliminary}
\label{sec:preliminary}
Graph Neural Networks are a powerful framework for learning representations on graph-structured data. Formally, let $\mathcal{G} = (\mathcal{V}, \mathcal{E})$ denote a graph with node set $\mathcal{V}$ and edge set $\mathcal{E}$. Each node $v \in \mathcal{V}$ is typically associated with an input feature vector $\mathbf{h}_v^{(0)} = \mathbf{x}_v$. The core principle of GNNs is the message passing mechanism, where each node iteratively updates its representation by aggregating information from its neighbors.

For a GNN with $K$ layers, the update rule for the representation of node $v$ at the $k$-th layer can be formulated as:
\begin{equation}
    \mathbf{h}_v^{(k)} = \text{COMBINE}^{(k)}\bigg[\mathbf{h}_v^{(k-1)}, \text{AGG}^{(k)}\left(\mathbf{h}_{u\in \mathcal{N}(v)}^{(k-1)}\right)\bigg]
\end{equation}
where $\mathbf{h}_v^{(k)}$ is the feature vector of node $v$ at layer $k$, $\mathcal{N}(v)$ is the set of immediate neighbors, $\text{AGG}^{(k)}(\cdot)$ is a permutation-invariant aggregation function like sum, mean, or max-pooling, and $\text{COMBINE}^{(k)}(\cdot)$ is an update function that fuses the node's previous representation with the aggregated neighborhood message.

After $K$ iterations of message passing, the final node representation $\mathbf{h}_v^{(K)}$ captures the structural information within its $K$-hop neighborhood and can be used for downstream tasks such as node classification or link prediction.

\section{Methodology}
\label{sec:method}
In this section, we present \textbf{CO-EVOLVE}, a cyclic, mutually reinforcing framework that unifies GNNs and LLMs to address robustness in heterophilous graph learning. Unlike static pipelines that treat one modality as fixed ground truth, CO-EVOLVE considers graph topology and semantic embeddings as latent variables that co-evolve via alternating optimization. As illustrated in Figure~\ref{fig:framework}, our framework consists of four integrated components: (1) \textbf{Structure-Conditioned Soft Prompting}, which encodes global topological properties into soft prompts to ground the LLM's reasoning and mitigate hallucinations; (2) \textbf{Adaptive Semantic-Guided Structure Learning}, which reconstructs the graph topology using evolved semantic representations to capture both homophily and structural complementarity; (3) a \textbf{Hard-Structure Conflict-Aware Loss}, which resolves dissonance between discordant semantic similarity and structural connectivity; and (4) \textbf{Uncertainty-Gated Consistency}, which stabilizes the bidirectional knowledge transfer by weighting supervision signals based on model confidence. The entire framework is trained using a Gauss-Seidel optimization strategy to minimize a unified objective function, ensuring simultaneous improvement of both modalities.

\subsection{Structure-Conditioned Soft Prompting}
Large Language Models possess immense semantic knowledge but suffer from structural blindness, leading to a Cold Start problem where the lack of topological awareness causes hallucinations on graph data \cite{wang2025graph,li2024glbench}. Existing integrations typically fail to address this by treating modalities separately or in rigid pipelines; LLM-first methods bake in semantic errors without structural guidance \cite{he2024harnessing,li2024zerog}, while GNN-first approaches lack continuous feedback to guide the LLM's reasoning process \cite{chen2024llaga}.

To bridge this gap, we introduce Structure-Conditioned Soft Prompting, a mechanism that injects global graph structural information into the LLM's latent space \emph{before} inference.

We first capture the high-order structural information of each node using a GNN encoder. Given the current dynamic graph adjacency matrix $\tilde{A}$ and node features $\mathbf{X}$, the GNN generates a structural embedding $\mathbf{H}_{struct}^{(i)}$ for each node $v_i$. This embedding condenses the node's local neighborhood and global topological role into a dense vector. To translate this structural signal into a format intelligible to the LLM, we employ a lightweight Multi-Layer Perceptron (MLP) as a projector. This projector maps the structural embedding $\mathbf{H}_{struct}^{(i)}$ into the continuous token embedding space of the LLM, creating a sequence of Soft Prompts $\mathbf{P}_{struct}^{(i)}$:
\begin{equation}
    \mathbf{P}_{struct}^{(i)} = \text{MLP}_{proj}(\mathbf{H}_{struct}^{(i)})
\end{equation}
Unlike discrete hard prompts, these soft prompts are differentiable vectors that can learn to encode complex, non-linguistic structural signals.

We then prepend these structure-derived prompts to the token embeddings of the node's raw textual attributes $\mathbf{T}_i$. The fused input sequence for node $v_i$ becomes:
\begin{equation}
    \text{Input}_i = [\mathbf{P}_{struct}^{(i)}, \text{Embedding}(\mathbf{T}_i)]
\end{equation}
The LLM, fine-tuned with Low-Rank Adaptation (LoRA), processes this structure-augmented sequence. By conditioning on $\mathbf{P}_{struct}^{(i)}$, the LLM's attention mechanism is grounded in the graph topology, allowing it to generate Structure-Aware Semantic Embeddings $\mathbf{H}_{sem}$ that reflect both textual content and structural context.

\begin{figure}[htbp]
    \centering
    \begin{subfigure}[t]{0.48\linewidth}
        \centering
        \includegraphics[width=\linewidth]{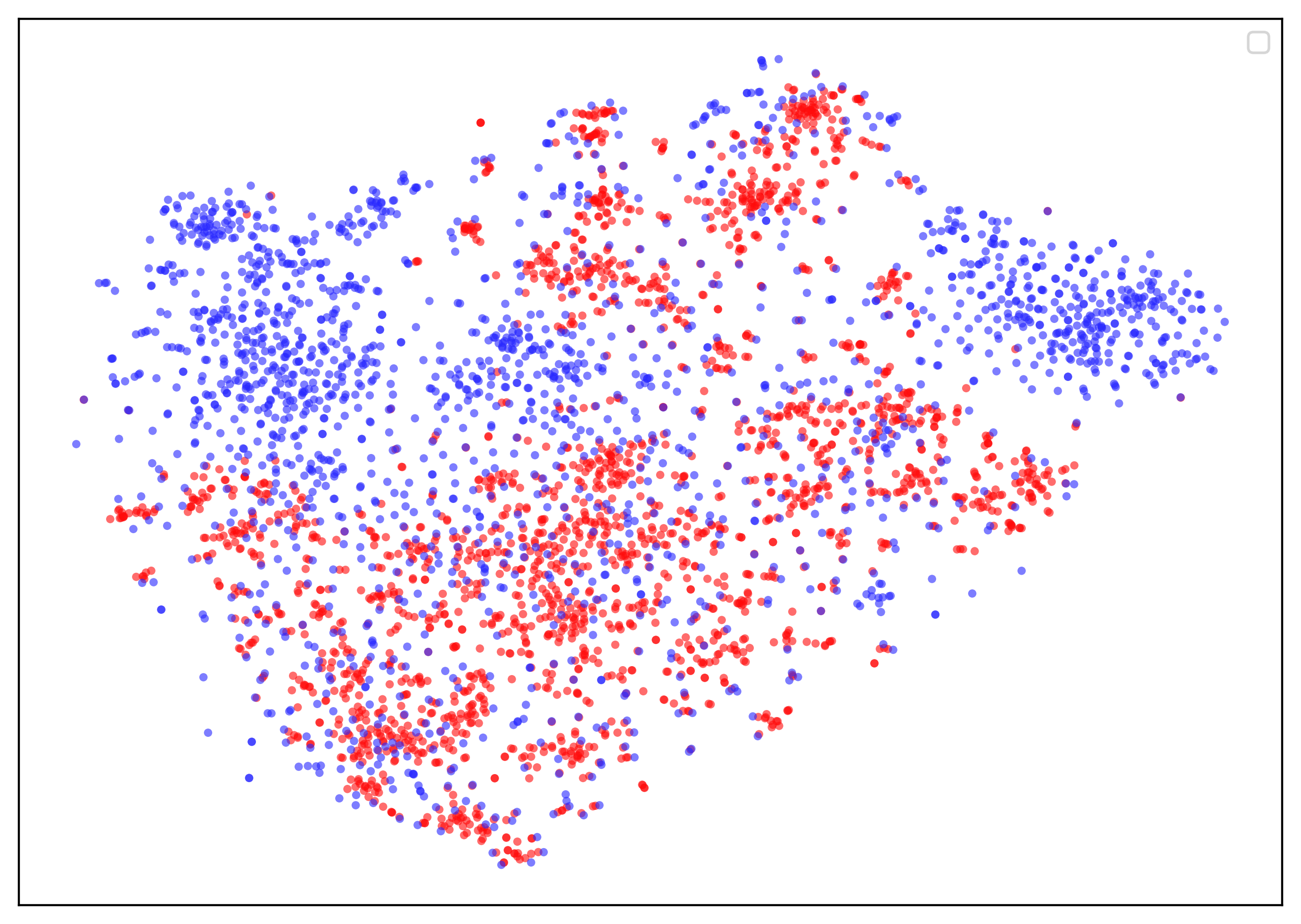}
        \caption{t-SNE without Soft Prompts}
    \end{subfigure}\hfill
    \begin{subfigure}[t]{0.48\linewidth}
        \centering
        \includegraphics[width=\linewidth]{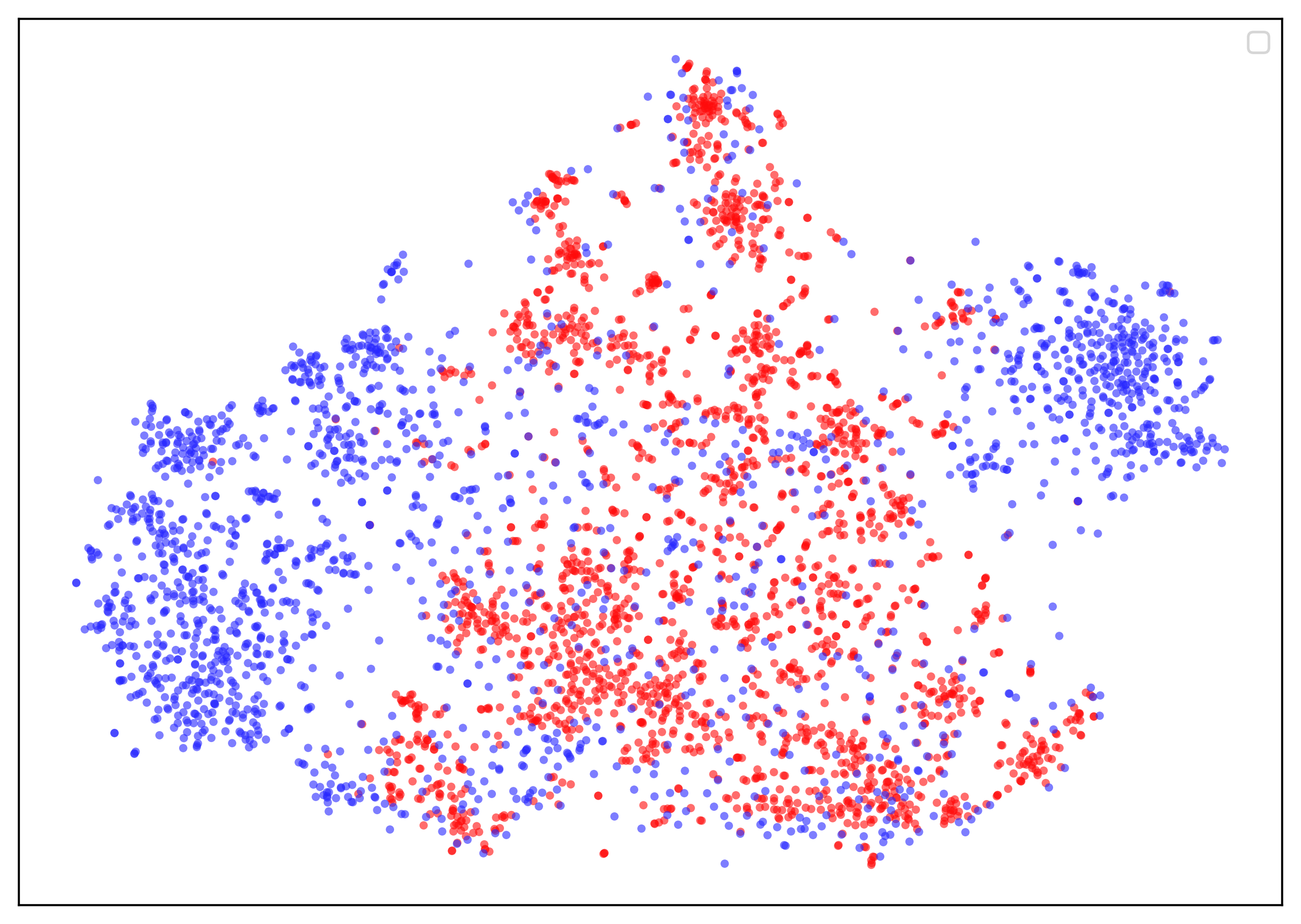}
        \caption{t-SNE with Soft Prompts}
    \end{subfigure}
    \caption{t-SNE visualization of node embeddings from the LLM with and without structure-conditioned soft prompts.}
    \label{fig:tsne_soft_prompts}
\end{figure}
To verify the effectiveness of injected structural prompts, we visualize node embeddings from the LLM both with and without structure-conditioned soft prompts using t-SNE (Figure \ref{fig:tsne_soft_prompts}). Without prompts, clusters are driven purely by textual similarity, often intermingled with overlapping class boundaries. In contrast, with soft prompts, clusters become much more distinct and well-separated into cohesive, class-aligned groups. This demonstrates that soft prompts successfully inject topological awareness, allowing the LLM to separate semantically similar but structurally distinct nodes.

\subsection{Adaptive Semantic-Guided Structure Learning}
Real-world graphs, often noisy and heterophilous, limit GNN performance when relying on static structures, as standard learning methods enforcing smoothness fail to capture complementary but varied relationships \cite{yang2025flag,wang2025graph}. Traditional Graph Structure Learning (GSL) approaches aggravate this by assuming homophily and using simple similarity metrics that introduce False Semantic Friends and miss Hidden Structural Neighbors \cite{jin2020graph}, while existing Graph-LLM frameworks often use static graphs that prevent the GNN from leveraging the LLM's evolving semantic insights \cite{yang2025flag,wang2025bridging}.

To overcome these limitations, we propose an Adaptive Semantic-Guided Structure Learning module. This component uses the LLM's updated semantic embeddings $\mathbf{H}_{sem}$ to reconstruct the graph topology dynamically. It is designed to capture complex, non-linear relationships beyond simple similarity and to adaptively balance trust between the static ground truth and the learned structure.

Instead of relying on a fixed metric like cosine similarity, we employ a Multi-Head Learnable Metric Function. This allows the model to capture diverse types of relationships, e.g., one head might capture homophily, another heterophily. For each head $k$, we compute the pairwise similarity between nodes $i$ and $j$ as:
\begin{equation}
    S_{ij}^{(k)} = (\mathbf{h}_i^{sem})^\top \mathbf{W}_k \mathbf{h}_j^{sem}
\end{equation}
where $\mathbf{W}_k$ is a learnable weight matrix for head $k$. The final semantic similarity matrix $\mathbf{S}$ is obtained by aggregating these heads, e.g., via averaging or concatenation followed by projection, enabling the graph to encode rich, multi-faceted structural patterns.

However, not all nodes require structure learning to the same extent. Some nodes have reliable connections in the static graph, while others are isolated or noisy. To handle this, we introduce a Node-Adaptive Gating Factor $\alpha_i \in [0, 1]$ that determines how much to trust the static structure versus the learned semantic graph. We compute $\alpha_i$ using a dedicated MLP:
\begin{equation}
    \alpha_i = \sigma(\text{MLP}_{gate}(\mathbf{h}_i^{sem}))
\end{equation}
A high $\alpha_i$ indicates the model trusts the static edges, while a low $\alpha_i$ signals a need to rely on the learned semantic connections. The fused adjacency candidate $A_{fused}$ is computed as:
\begin{equation}
    A_{fused}^{(ij)} = \alpha_i A_{static}^{(ij)} + (1-\alpha_i) S_{ij}
\end{equation}
This mechanism allows the model to rewire the graph locally where needed while preserving robust static structures.

To maintain computational efficiency and filter out noise from weak correlations, we apply a sparsification step to the fused matrix. We retain only the top-$k$ strongest connections for each node:
\begin{equation}
    \tilde{A} = \text{TopK}(A_{fused})
\end{equation}
This results in a sparse, high-quality dynamic graph $\tilde{A}$ that is fed into the GNN for the next iteration of training. This evolved structure provides a corrected topological view that complements the LLM's semantic view.

\begin{figure}[htbp]
    \centering
    \begin{subfigure}[t]{0.48\linewidth}
        \centering
        \includegraphics[width=\linewidth]{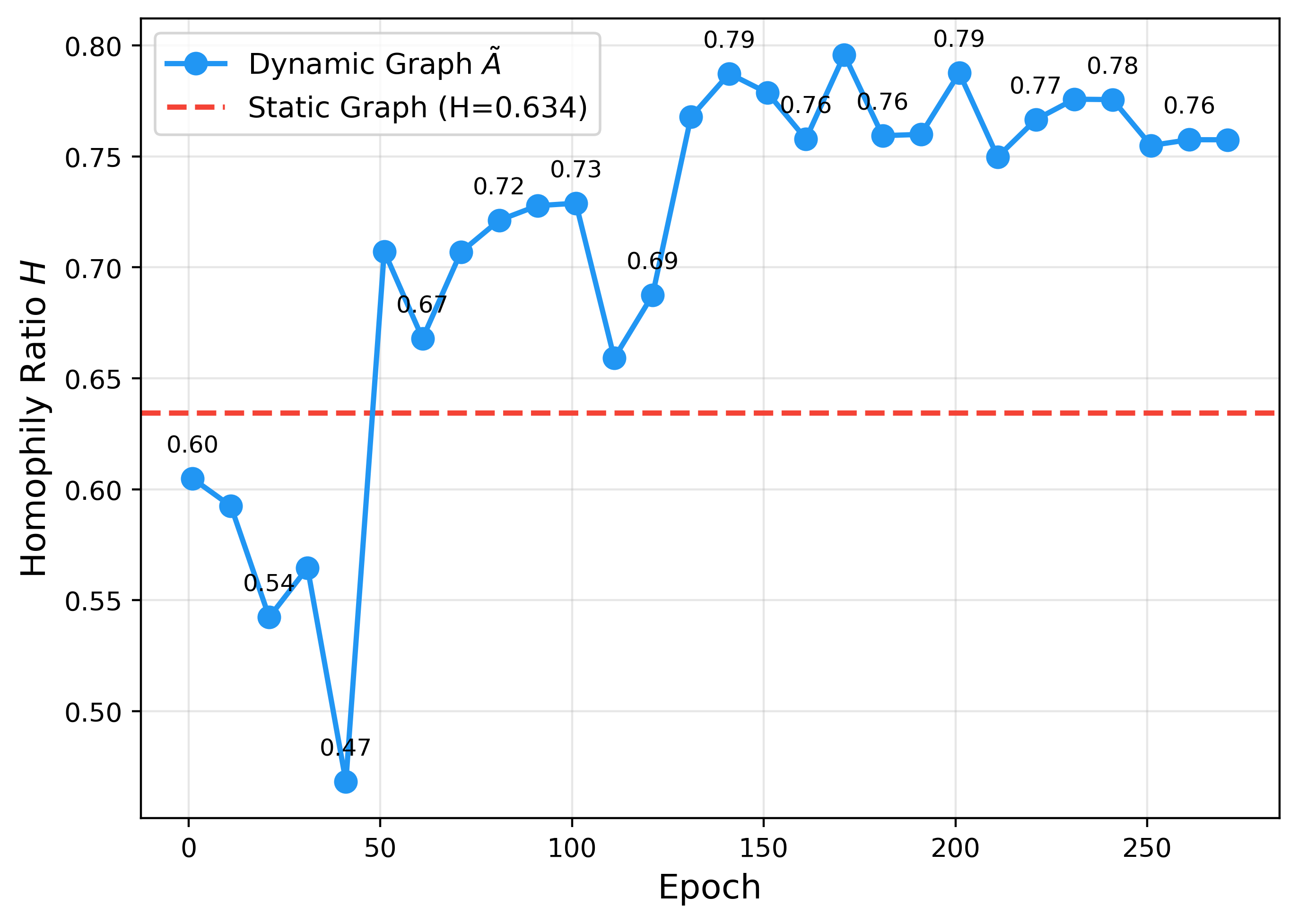}
        \caption{Homophily Ratio Evolution}
        \label{fig:homophily_evolution}
    \end{subfigure}\hfill
    \begin{subfigure}[t]{0.48\linewidth}
        \centering
        \includegraphics[width=\linewidth]{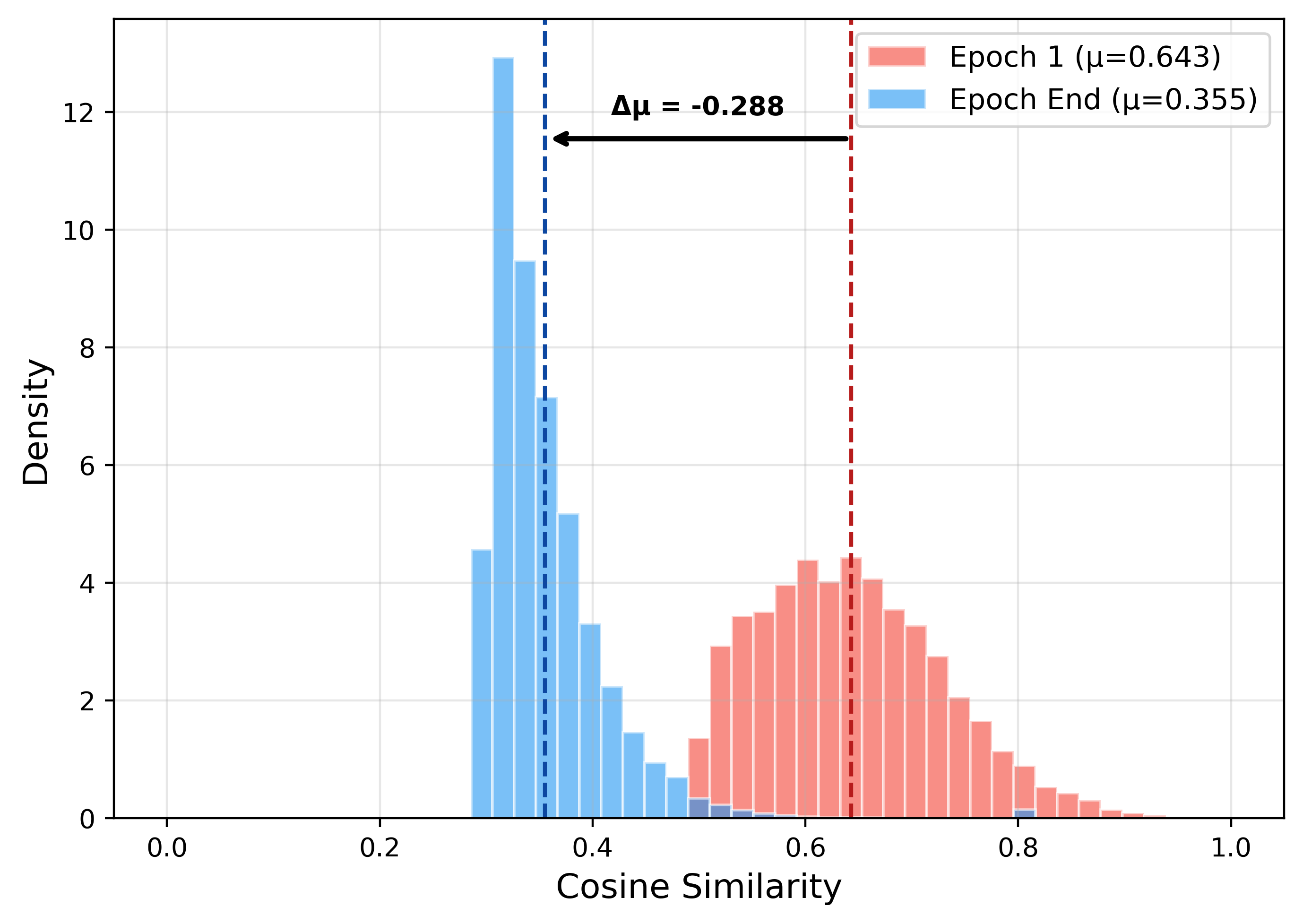}
        \caption{Conflict Loss Distribution}
        \label{fig:conflict_loss}
    \end{subfigure}
    \caption{Homophily Ratio Evolution and Conflict Loss Distribution.}
    \label{fig:homophily_evolution_conflict_loss}
\end{figure}

To validate the learned structure's quality, we track the Homophily Ratio ($H$) of the dynamic graph $\tilde{A}$ across training epochs (Figure \ref{fig:homophily_evolution}). The homophily ratio measures the fraction of edges that connect nodes of the same class, and is formalized as:
\begin{equation}
    H = \frac{\sum_{i,j \in \mathcal{E}} \mathbf{1}[y_i = y_j]}{|\mathcal{E}|}
\end{equation}

We observe that, although the homophily ratio of the dynamic graph slightly fluctuates initially, it rapidly increases and surpasses the static graph's baseline ($0.634$), maintaining a consistently high level of approximately $0.75$ to $0.79$ throughout the later epochs. This indicates that the proposed structure learning effectively filters out heterophilous noise and recovers missing homophilous edges, creating a cleaner graph structure for GNN propagations.

\subsection{Hard-Structure Conflict-Aware Loss}
Heterophilous graphs often exhibit Semantic-Structural Dissonance \cite{zheng2025enhancing}, where nodes can be False Semantic Friends or Hidden Structural Neighbors, causing standard contrastive losses to merge distinct classes based on superficial textual similarity. Existing methods \cite{wang2025graph} fail to explicitly model this conflict, treating all neighbors as positives and averaging out contradictory signals from the LLM and graph structure, which leads to suboptimal decision boundaries that do not respect true topological constraints.

To rigorously resolve this dissonance, we introduce a Hard-Structure Conflict-Aware Loss. This objective function uses global graph diffusion to explicitly penalize semantic similarities that contradict structural reality and reinforce structural connections that are semantically weak.

To capture high-order structural proximity robust to local noise, we employ the Personalized PageRank (PPR) matrix $\mathbf{\Pi}$ as a global topological ground truth. Unlike direct adjacency, which is sparse and sensitive to missing edges, PPR models the probability that a random walker starting at node $i$ will visit node $j$ in the steady state, thereby revealing implicit community structures and long-range dependencies. The closed-form solution of PPR is given by:
\begin{equation}
    \mathbf{\Pi} = \gamma (\mathbf{I} - (1-\gamma) \hat{\mathbf{A}})^{-1}
\end{equation}
where $\gamma \in (0, 1)$ is the restart probability controlling the diffusion range, and $\hat{\mathbf{A}} = \tilde{\mathbf{D}}^{-1/2} \tilde{\mathbf{A}} \tilde{\mathbf{D}}^{-1/2}$ is the normalized adjacency matrix with self-loops. This dense matrix $\mathbf{\Pi}$ provides a smoothed, global view of structural affinity that guides the alignment of semantic embeddings.

We dynamically identify two critical sets of pairs for each node $v_i$:
Hard Conflict Negatives ($\mathcal{H}_i$) and Structural Positives ($\mathcal{P}_i$). For $\mathcal{H}_i$, pairs that are semantically similar ($\mathbf{z}_i^\top \mathbf{z}_k > \tau$) but structurally disconnected ($\mathbf{\Pi}_{ik} < \epsilon$) are added. These represent False Semantic Friends that the model must learn to push apart. For $\mathcal{P}_i$, pairs with high global structural diffusion scores ($\mathbf{\Pi}_{ij} > \alpha$) are selected, representing True Structural Neighbors that should be close in the embedding space.

We minimize a margin-based contrastive loss on these specific sets to warp the semantic manifold to respect topological boundaries:
\begin{equation}
    \begin{split}
        \mathcal{L}_{conflict} = \frac{1}{N} \sum_{i=1}^{N} \bigg[ &\sum_{j \in \mathcal{P}_i} \max(0, \Delta_+ - \mathbf{z}_i^\top \mathbf{z}_j) + \\ & \lambda \sum_{k \in \mathcal{H}_i} \max(0, \mathbf{z}_i^\top \mathbf{z}_k - \Delta_-) \bigg]
    \end{split}
\end{equation}
where $\mathbf{z}$ are the normalized embeddings, $\Delta_+$ and $\Delta_-$ are margins for positives and negatives, and $\lambda$ balances the penalty. This loss forces the embedding space to conform to the high-order structural constraints of the graph, effectively resolving heterophilous conflicts.

To analyze the resolution of semantic-structural dissonance, we track the cosine similarity of hard negative pairs. As shown in Figure \ref{fig:conflict_loss}, at Epoch 1, the distribution is skewed towards high similarity ($0.50-0.90$ with $\mu=0.643$). By the final epoch, it shifts significantly to lower similarity ($0.30-0.50$ with $\mu=0.355$). This confirms that the loss function effectively warps the semantic space to separate structurally disconnected nodes, resolving the conflict between misleading textual similarity and topological reality.

\subsection{Uncertainty-Gated Consistency and Fusion}
Standard knowledge alignment methods assume a correct teacher \cite{yu2025samgpt,fang2026knowledge}, but in our co-evolution framework, blindly aligning evolving and potentially erroneous models propagates noise and degrades robustness. Most co-training approaches lack reliability assessment, using symmetric consistency losses indiscriminately \cite{yang2025flag,jin2023patton}. Consequently, unchecked errors flow between the LLM and GNN, leading to a ``race to the bottom'' rather than mutual enhancement.

\begin{figure}[htbp]
    \centering
    \includegraphics[width=0.6\linewidth]{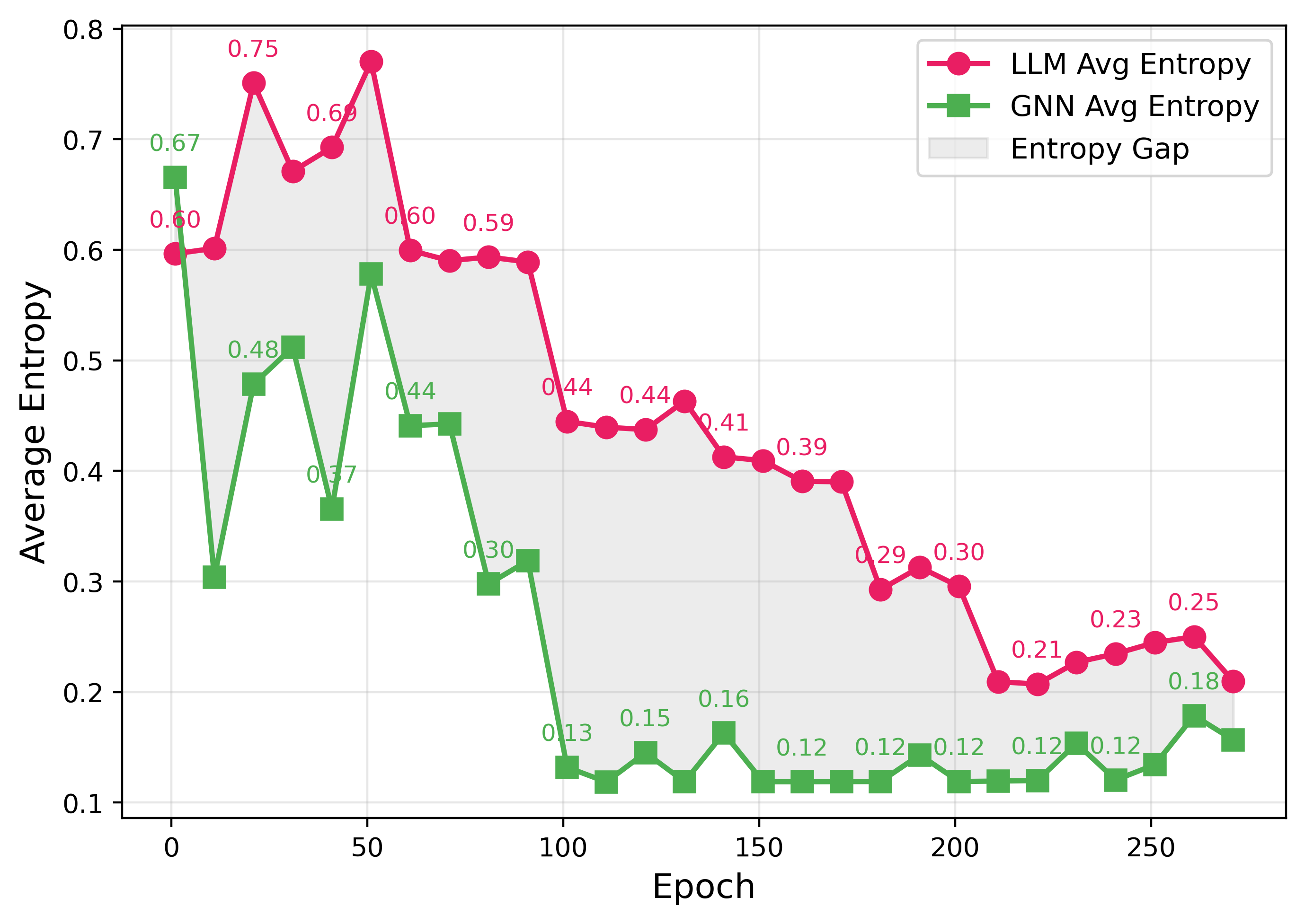}
    \caption{Entropy Evolution of LLM and GNN.}
    \label{fig:uncertainty_curve}
\end{figure}
To prevent this, we propose an Uncertainty-Gated Consistency mechanism. This approach ensures that information flows primarily from the more confident view to the less confident one, acting as a noise filter during co-evolution.

We first quantify the uncertainty of each model (LLM and GNN) for every node using the normalized entropy of their predicted probability distributions $P$:
\begin{equation}
    \mathbb{H}(P) = -\sum_{c=1}^{C} P_c \log P_c
\end{equation}
A high entropy indicates the model is uncertain about its prediction, while low entropy suggests high confidence.

We then modulate the bidirectional alignment loss (KL divergence) using the Teacher's Confidence ($1 - \text{Entropy}$). The LLM updates its parameters to match the GNN's distribution only when the GNN is confident, and vice versa:
\begin{equation}
    \begin{split}
        \mathcal{L}_{cons} = &(1 - \mathbb{H}(P_{GNN})) \cdot \text{KL}(P_{GNN} || P_{LLM}) +\\ & (1 - \mathbb{H}(P_{LLM})) \cdot \text{KL}(P_{LLM} || P_{GNN})
    \end{split}
\end{equation}
This gating mechanism effectively amplifies high-quality supervision signals while suppressing noisy ones, stabilizing the co-evolutionary process.

Finally, at prediction time, we obtain final predictions by fusing the outputs of both models. Instead of a simple average, we use a learnable gating network $\beta$ that dynamically weighs the two views. Crucially, this gate conditions not just on uncertainty, but also on the Structural Embedding $\mathbf{H}_{struct}$, allowing the fusion strategy to adapt to the node's topological context, e.g., relying more on GNNs for well-connected nodes and LLMs for isolated ones:
\begin{equation}
    \beta = \sigma(\text{MLP}([\mathbb{H}(P_{LLM}), \mathbb{H}(P_{GNN}), \mathbf{H}_{struct}]))
\end{equation}
The final prediction is given by:
\begin{equation}
    Y_{final} = \beta P_{LLM} + (1-\beta) P_{GNN}
\end{equation}

To demonstrate the stability of co-evolution, we track the average Entropy of LLM and GNN predictions over training epochs. We find from Figure \ref{fig:uncertainty_curve} that the average entropy of both the LLM and GNN progressively decreases over the training epochs, converging to much lower uncertainty levels, while the entropy gap between the two models narrows significantly. This confirms that gating consistency based on uncertainty ensures a stable mutual enhancement, where the more confident view progressively guides the weaker one, preventing the amplification of errors.

\subsection{Training Strategy and Optimization}
To overcome the non-differentiability of Top-k operation and the instability and memory bottlenecks of joint optimization methods, we employ a Gauss-Seidel Alternating Optimization strategy. This decouples the training into two alternating phases, allowing each modality to learn from the most up-to-date version of the other. The training cycle proceeds as follows:

\noindent\textbf{1. Warm-up:} We pre-train the GNN and LLM independently for a few epochs to initialize their parameters and avoid minimalizing the consistency loss through trivial solutions.

\noindent\textbf{2. Step A (LLM Update):} We freeze the GNN parameters. The LLM receives structural soft prompts generated by the current GNN and updates its LoRA parameters. The objective minimizes the supervised task loss, i.e., the cross-entropy loss, the conflict-aware contrastive loss, and the uncertainty-gated consistency loss:
\begin{equation}
    \mathcal{L}_{total} = \mathcal{L}_{task} + \mathcal{L}_{conflict} + \mathcal{L}_{cons}
\end{equation}

\noindent\textbf{3. Step B (Graph Re-Construction):} Using the updated LLM, we generate fresh semantic embeddings $\mathbf{H}_{sem}$ and construct the new dynamic graph $\tilde{A}$ via the Adaptive Semantic-Guided Structure Learning module. This refreshed topology provides a corrected structural view for the GNN.

\noindent\textbf{4. Step C (GNN Update):} We freeze the LLM parameters. The GNN takes the new graph $\tilde{A}$ as input and updates its weights. The objective is analogous to Step A.

This cycle repeats until convergence. By iteratively refining the graph structure via the LLM and the semantic prompts via the GNN, CO-EVOLVE ensures that both views progressively align and enhance each other. Algorithm \ref{alg:framework} summarizes the training procedure.

\subsection{Inference Phase}
\begin{figure}[!h]
    \centering
    \includegraphics[width=0.35\textwidth]{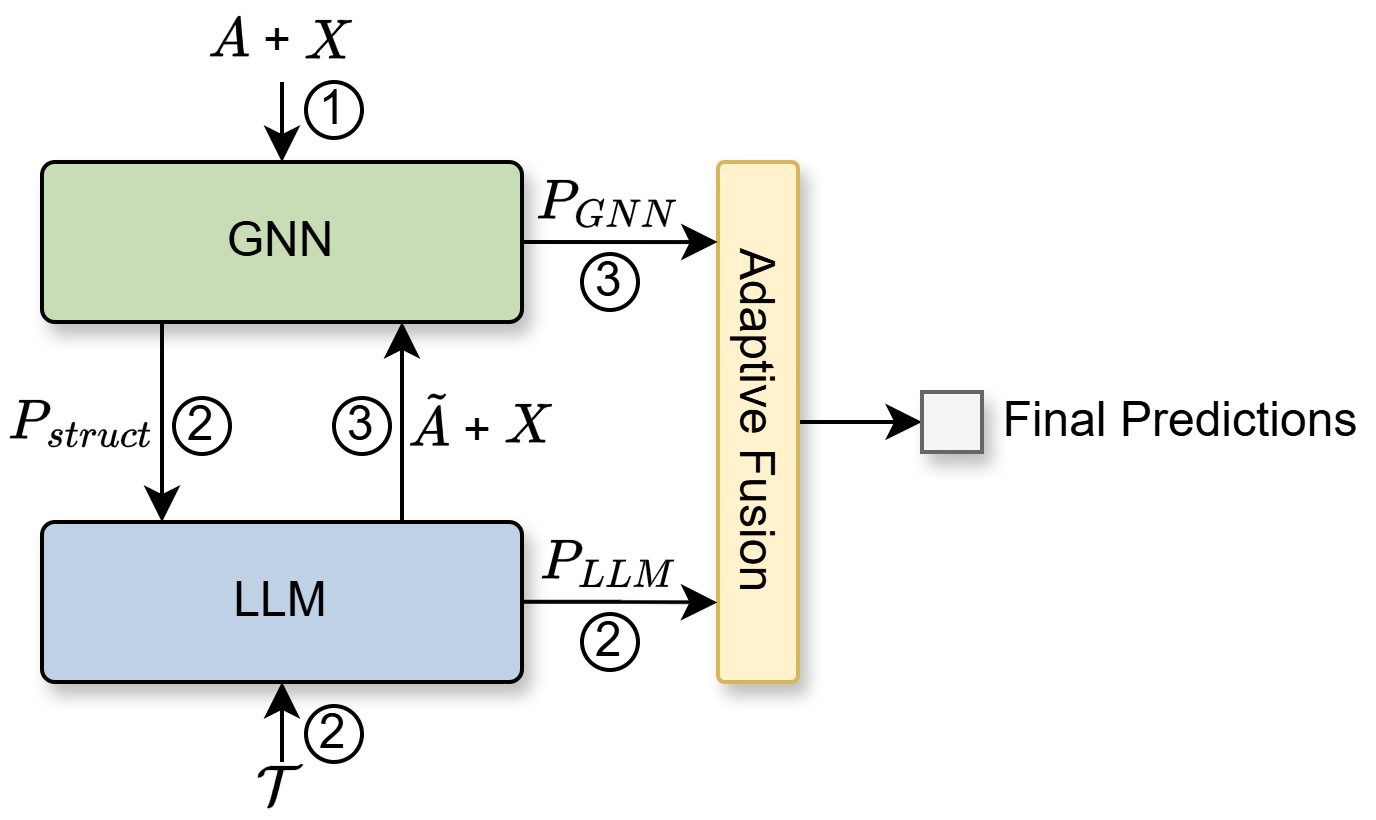}
    \caption{Overview of the multi-stage sequential inference protocol in CO-EVOLVE. The process involves structural context extraction via GNN, LLM semantic reasoning via soft prompts, dynamic graph reconstruction, and a final uncertainty-gated fusion of both views.}
    \label{fig:inference}
\end{figure}
Unlike standard single-pass inference, we employ a multi-stage sequential inference protocol, as shown in Figure \ref{fig:inference}, to fully leverage the co-evolutionary mechanism trained during the optimization phase. First, the GNN extracts structural context from the static graph, which is projected into soft prompts to guide the LLM's semantic reasoning. The resulting semantic embeddings from the LLM are used to reconstruct a refined dynamic graph, filtering noise and recovering missing edges. A second GNN pass is executed on this optimized topology. Finally, we fuse the predictions from both views using an uncertainty-gated mechanism that weighs each model based on its local confidence and structural context.

\section{Experiment}
\label{sec:experiment}
In this section, we comprehensively evaluate the performance of our proposed framework CO-EVOLVE on multiple benchmarks. Specifically, we designed a series of experiments to answer the following research questions:
\textbf{Q1:} How does CO-EVOLVE perform on standard node classification benchmarks compared to state-of-the-art baselines?
\textbf{Q2:} In real-world environments where misleading textual descriptions, specifically false semantic friends, exist, how robust is CO-EVOLVE compared to existing methods?
\textbf{Q3:} In sparse networks where critical structural links are missing, can CO-EVOLVE effectively maintain performance?
\textbf{Q4:} How does each individual component of CO-EVOLVE contribute to its overall performance?
\textbf{Q5:} What is the impact of hyper-parameters on the final performance of CO-EVOLVE?

\subsection{Experimental Setup}
\subsubsection{Datasets}
To evaluate the effectiveness of CO-EVOLVE, we conduct our experiments on three real-world datasets that contain both graph structure and text attributes: Reddit \cite{li2024glbench}, Instagram \cite{li2024glbench}, and WikiCS \cite{liu2023one}. Specifically, Reddit is a user social network where edges represent replies, and the task is to distinguish between popular and normal users based on node features consisting of up to three recent subreddit posts. Instagram is a social network where edges reflect following relationships, and the task classifies accounts as commercial or normal using node features that contain the user's personal introduction. Finally, WikiCS is a Wikipedia reference network linking connected pages, where nodes are classified into computer science categories using features derived from article names and content. Detailed dataset statistics are shown in Appendix \ref{tab:appendix_dataset_statistics}.

\subsubsection{Baselines and Implementations}
To demonstrate the superiority of CO-EVOLVE, we compare it against 10 competitive baselines across three categories. GNN-only models: GCN \cite{kipf2017semi}, GAT \cite{velickovic2018graph}. LM-only models: Sent-BERT \cite{reimers2019sentence}, RoBERTa \cite{liu2019roberta}. Graph-LLM Models: TAPE \cite{he2023harnessing}, ZeroG \cite{li2024zerog}, LLaGA \cite{chen2024llaga}, InstructGLM \cite{ye2024language}, FLAG \cite{yang2025flag}, and Patton \cite{jin2023patton}.

Our CO-EVOLVE framework is implemented using PyTorch and PyTorch Geometric. For the semantic view, we employ Llama-3.2-1B \cite{grattafiori2024llama} as the base LLM, combined with LoRA \cite{hu2022lora} parameter-efficient fine-tuning with rank $r=4$. The structural view relies on a 2-layer GCN with a hidden dimension of 128 and an output structure dimension of 64. The bidirectional framework is trained via AdamW optimizers: the LLM incorporates a learning rate of $1\times 10^{-4}$ and the GNN uses $5\times 10^{-3}$. We use a cosine learning rate scheduler with a 10\% step warm-up profile. In the Semantic-Guided Structure Learning module, the pruning threshold $k$ is set to 10 following \cite{yang2025flag}. For the Conflict-Aware Loss, we set the structural positive threshold $\alpha = 0.7$, the structural irrelevance threshold $\epsilon = 0.3$, and the semantic hallucination threshold $\tau = 0.5$. The experiments are conducted on NVIDIA A100 GPUs with 80GB memory.

\subsection{Benchmarks (Q1)}
We first analyze the performance of all methods on three benchmarks as shown in Table~\ref{tab:baselines}. CO-EVOLVE demonstrates strong and highly competitive performance across all datasets. On the Instagram dataset, CO-EVOLVE achieves the highest Accuracy 69.74\%, significantly outperforming modern Graph-LLM baselines like FLAG and ZeroG. On WikiCS, CO-EVOLVE yields the best Accuracy 85.35\% and the second-best F1 81.89\%, surpassing TAPE and FLAG. On the Reddit dataset, our method ranks second in both Accuracy and F1, marginally behind LLaGA while broadly outperforming all other baselines. The robust performance across structurally and semantically diverse datasets highlights the capability of the bidirectional co-evolution framework to optimally fuse both modalities, thereby preventing unidirectional error propagation.

\begin{table}[htbp]
    \centering
    \caption{Performance comparison of different methods on Reddit, Instagram, and Wikics datasets.}
    \begin{tabular}{lcccccc}
        \toprule
        \multicolumn{1}{l}{\multirow{2}[4]{*}{Methods}} & \multicolumn{2}{c}{Reddit}          & \multicolumn{2}{c}{Instagram}       & \multicolumn{2}{c}{Wikics}                                                                                                                            \\
        \cmidrule(lr){2-3} \cmidrule(lr){4-5} \cmidrule(lr){6-7}
                                                        & Acc                                 & F1                                  & \multicolumn{1}{c}{Acc}             & \multicolumn{1}{c}{F1}              & \multicolumn{1}{c}{Acc}             & \multicolumn{1}{c}{F1}              \\
        \midrule
        GCN                                             & 53.87                               & 53.17                               & 63.96                               & 53.36                               & 76.49                               & 73.42                               \\
        GAT                                             & 52.65                               & 52.46                               & 63.72                               & 54.95                               & 76.83                               & 75.72                               \\
        Sent-BERT                                       & 55.99                               & 55.77                               & 61.04                               & 54.79                               & 74.20                               & 72.27                               \\
        RoBERTa                                         & 55.85                               & 55.78                               & 58.56                               & 53.11                               & 74.77                               & 74.49                               \\
        TAPE                                            & 59.41                               & 59.18                               & 66.80                               & 48.40                               & \textcolor{blue}{\underline{84.44}} & \textcolor{red}{\textbf{82.47}}     \\
        ZeroG                                           & 61.88                               & 58.02                               & \textcolor{blue}{\underline{68.51}} & \textcolor{blue}{\underline{58.35}} & 78.82                               & 77.32                               \\
        LLaGA                                           & \textcolor{red}{\textbf{62.87}}     & \textcolor{red}{\textbf{61.86}}     & 64.15                               & 52.46                               & 71.64                               & 68.91                               \\
        InstructGLM                                     & 54.46                               & 51.92                               & 60.30                               & 53.14                               & 74.40                               & 73.55                               \\
        FLAG                                            & 58.29                               & 58.09                               & 68.05                               & \textcolor{red}{\textbf{58.48}}     & 83.04                               & 78.90                               \\
        Patton                                          & 58.11                               & 56.53                               & 64.90                               & 55.25                               & 82.05                               & 78.16                               \\
        CO-EVOLVE                                       & \textcolor{blue}{\underline{61.97}} & \textcolor{blue}{\underline{60.84}} & \textcolor{red}{\textbf{69.74}}     & 57.11                               & \textcolor{red}{\textbf{85.35}}     & \textcolor{blue}{\underline{81.89}} \\
        \bottomrule
    \end{tabular}%
    \label{tab:baselines}%
\end{table}%

\subsection{False Semantic Friends (Q2)}
While public benchmarks are relatively clean, real-world graphs frequently suffer from noisy or deceptive attribute information such as spam accounts with misleading text \cite{yang2025flag}. These deceptive attributes severely degrade the performance of existing baselines: LM-centric models are directly misled to generate erroneous embeddings that incorrectly group cross-class nodes together, while GNN-centric methods inevitably propagate these corrupted semantic features across structural neighborhoods. To rigorously quantify the CO-EVOLVE's resilience against such false semantic friends, we construct adversarial scenarios by deliberately planting misleading textual signals. Specifically, we identify structurally distant, cross-class node pairs using the PPR matrix calculations where $\Pi_{ij} < 0.3$. For a randomly selected $r \in (10\%, 20\%, 30\%)$ of these pairs $(v_i, v_j)$, we swap the text attributes of $v_i$ with text randomly selected from the class of $v_j$. This text perturbation strictly isolates the attack to the semantic modality without modifying the true graph topology. This stress test fundamentally evaluates our model's ability to suppress semantic hallucinations compared to vulnerable baselines.

\begin{table}[htbp]
    \centering
    \caption{Performance comparison of different methods on Reddit, Instagram, and Wikics datasets under 10\% semantic corruption.}
    \begin{tabular}{lcccccc}
        \toprule
        \multicolumn{1}{l}{\multirow{2}[4]{*}{Methods}} & \multicolumn{2}{c}{Reddit}          & \multicolumn{2}{c}{Instagram}       & \multicolumn{2}{c}{Wikics}                                                                                                                            \\
        \cmidrule(lr){2-3} \cmidrule(lr){4-5} \cmidrule(lr){6-7}
                                                        & Acc                                 & F1                                  & \multicolumn{1}{c}{Acc}             & \multicolumn{1}{c}{F1}              & \multicolumn{1}{c}{Acc}             & \multicolumn{1}{c}{F1}              \\
        \midrule
        GCN                                             & 47.92                               & 46.31                               & 56.92                               & 49.41                               & 68.12                               & 65.91                               \\
        GAT                                             & 46.78                               & 46.62                               & 56.67                               & 48.82                               & 68.32                               & 68.19                               \\
        Sent-BERT                                       & 51.40                               & 50.19                               & 54.92                               & 49.31                               & 66.76                               & 65.05                               \\
        RoBERTa                                         & 50.25                               & 50.20                               & 52.70                               & 47.76                               & 68.31                               & 67.04                               \\
        TAPE                                            & 54.91                               & 54.67                               & 61.81                               & 45.73                               & 77.09                               & \textcolor{blue}{\underline{75.98}} \\
        ZeroG                                           & \textcolor{blue}{\underline{56.59}} & 55.09                               & 62.59                               & 53.30                               & 72.07                               & 70.50                               \\
        LLaGA                                           & 56.19                               & \textcolor{blue}{\underline{55.33}} & 57.32                               & 46.92                               & 64.08                               & 62.28                               \\
        InstructGLM                                     & 49.59                               & 47.28                               & 54.90                               & 49.40                               & 67.78                               & 66.49                               \\
        FLAG                                            & 55.50                               & 53.31                               & \textcolor{blue}{\underline{64.82}} & \textcolor{blue}{\underline{54.60}} & \textcolor{blue}{\underline{79.13}} & 75.61                               \\
        Patton                                          & 54.42                               & 52.89                               & 60.67                               & 52.73                               & 78.83                               & 72.80                               \\
        CO-EVOLVE                                       & \textcolor{red}{\textbf{61.51}}     & \textcolor{red}{\textbf{59.32}}     & \textcolor{red}{\textbf{66.19}}     & \textcolor{red}{\textbf{56.61}}     & \textcolor{red}{\textbf{82.78}}     & \textcolor{red}{\textbf{81.19}}     \\
        \bottomrule
    \end{tabular}%
    \label{tab:baselines_fsf_10}%
\end{table}%

As observed in Tables~\ref{tab:baselines_fsf_10}, \ref{tab:baselines_fsf_20}, and \ref{tab:baselines_fsf_30}, CO-EVOLVE exhibits exceptional resilience against semantic corruption. Even at a severe 30\% perturbation rate, CO-EVOLVE experiences only an 8.75\% and 6.88\% absolute accuracy drop on the Reddit and WikiCS datasets, respectively. In stark contrast, LM-centric and Graph-LLM baselines like LLaGA suffer catastrophic degradation, e.g., a 23.36\% drop on Reddit, while GNNs like GCN face compounding neighbor-aggregation errors leading to over a 22\% accuracy drop.

\begin{table}[htbp]
    \centering
    \caption{Performance comparison of different methods on Reddit, Instagram, and Wikics datasets under 20\% semantic corruption.}
    \begin{tabular}{lcccccc}
        \toprule
        \multicolumn{1}{l}{\multirow{2}[4]{*}{Methods}} & \multicolumn{2}{c}{Reddit}          & \multicolumn{2}{c}{Instagram}       & \multicolumn{2}{c}{Wikics}                                                                                                                            \\
        \cmidrule(lr){2-3} \cmidrule(lr){4-5} \cmidrule(lr){6-7}
                                                        & Acc                                 & F1                                  & \multicolumn{1}{c}{Acc}             & \multicolumn{1}{c}{F1}              & \multicolumn{1}{c}{Acc}             & \multicolumn{1}{c}{F1}              \\
        \midrule
        GCN                                             & 42.92                               & 40.42                               & 48.61                               & 42.47                               & 58.18                               & 55.13                               \\
        GAT                                             & 39.99                               & 38.78                               & 47.42                               & 41.68                               & 58.43                               & 56.47                               \\
        Sent-BERT                                       & 46.80                               & 45.62                               & 46.81                               & 43.82                               & 58.38                               & 57.80                               \\
        RoBERTa                                         & 45.66                               & 44.60                               & 49.86                               & 42.45                               & 59.80                               & 58.57                               \\
        TAPE                                            & \textcolor{blue}{\underline{51.40}} & 49.23                               & 55.60                               & 42.28                               & 68.07                               & 59.30                               \\
        ZeroG                                           & 50.08                               & 46.99                               & 55.39                               & \textcolor{blue}{\underline{51.73}} & 63.78                               & 62.12                               \\
        LLaGA                                           & 48.38                               & 47.61                               & 49.42                               & 40.39                               & 55.32                               & 53.52                               \\
        InstructGLM                                     & 43.33                               & 41.29                               & 47.90                               & 42.20                               & 59.19                               & 57.82                               \\
        FLAG                                            & 50.60                               & \textcolor{blue}{\underline{49.39}} & \textcolor{blue}{\underline{60.29}} & 47.19                               & \textcolor{blue}{\underline{73.53}} & \textcolor{blue}{\underline{68.88}} \\
        Patton                                          & 49.22                               & 47.83                               & 54.98                               & 46.81                               & 69.51                               & 65.52                               \\
        CO-EVOLVE                                       & \textcolor{red}{\textbf{58.70}}     & \textcolor{red}{\textbf{57.42}}     & \textcolor{red}{\textbf{65.17}}     & \textcolor{red}{\textbf{53.81}}     & \textcolor{red}{\textbf{80.57}}     & \textcolor{red}{\textbf{77.84}}     \\
        \bottomrule
    \end{tabular}%
    \label{tab:baselines_fsf_20}%
\end{table}%

\begin{table}[htbp]
    \centering
    \caption{Performance comparison of different methods on Reddit, Instagram, and Wikics datasets under 30\% semantic corruption.}
    \begin{tabular}{lcccccc}
        \toprule
        \multicolumn{1}{l}{\multirow{2}[4]{*}{Methods}} & \multicolumn{2}{c}{Reddit}          & \multicolumn{2}{c}{Instagram}       & \multicolumn{2}{c}{Wikics}                                                                                                                            \\
        \cmidrule(lr){2-3} \cmidrule(lr){4-5} \cmidrule(lr){6-7}
                                                        & Acc                                 & F1                                  & \multicolumn{1}{c}{Acc}             & \multicolumn{1}{c}{F1}              & \multicolumn{1}{c}{Acc}             & \multicolumn{1}{c}{F1}              \\
        \midrule
        GCN                                             & 31.78                               & 30.40                               & 37.81                               & 34.50                               & 45.27                               & 43.20                               \\
        GAT                                             & 31.12                               & 29.88                               & 35.72                               & 32.50                               & 45.52                               & 43.08                               \\
        Sent-BERT                                       & 39.22                               & 39.04                               & 42.72                               & 38.33                               & 52.95                               & 50.59                               \\
        RoBERTa                                         & 39.68                               & 39.02                               & 41.02                               & 37.17                               & 54.33                               & 52.14                               \\
        TAPE                                            & 43.10                               & \textcolor{blue}{\underline{42.88}} & 48.37                               & 36.13                               & 61.20                               & 58.30                               \\
        ZeroG                                           & 43.07                               & 41.39                               & 47.62                               & 43.48                               & 55.88                               & 53.00                               \\
        LLaGA                                           & 39.51                               & 35.82                               & 49.31                               & 32.98                               & 45.09                               & 42.82                               \\
        InstructGLM                                     & 36.32                               & 35.62                               & 40.11                               & 35.32                               & 49.57                               & 48.11                               \\
        FLAG                                            & \textcolor{blue}{\underline{43.51}} & 42.29                               & \textcolor{blue}{\underline{54.23}} & \textcolor{blue}{\underline{46.41}} & \textcolor{blue}{\underline{66.21}} & \textcolor{blue}{\underline{63.42}} \\
        Patton                                          & 42.47                               & 42.31                               & 47.51                               & 40.43                               & 61.12                               & 56.38                               \\
        CO-EVOLVE                                       & \textcolor{red}{\textbf{53.22}}     & \textcolor{red}{\textbf{51.78}}     & \textcolor{red}{\textbf{61.49}}     & \textcolor{red}{\textbf{51.41}}     & \textcolor{red}{\textbf{78.47}}     & \textcolor{red}{\textbf{74.52}}     \\
        \bottomrule
    \end{tabular}%
    \label{tab:baselines_fsf_30}%
\end{table}%

This robust behavior stems directly from our Conflict-Aware Loss. When deceptive textual anomalies prompt the semantic view to falsely group the perturbed nodes with incorrect classes, the structurally grounded conflict condition accurately flags this semantic-structural dissonance. Subsequently, the hard-conflict loss explicitly penalizes these false similarities, precluding the Semantic Structure Learner from drawing spurious edges and isolating the corrupted representations before they propagate through the structure.

Ultimately, CO-EVOLVE successfully avoids the fatal pitfall of unidirectional error propagation from compromised semantic inputs, maintaining state-of-the-art performance in inherently noisy real-world graph environments.

\subsection{Missing Structural Links (Q3)}
In applications like citation networks or biological diagrams, the observed structure is often incomplete due to missing edges between truly related entities \cite{zheng2025enhancing}. This structural sparsity reveals a critical vulnerability in standard baselines: GNN-centric models natively underperform due to broken message-passing pathways that prevent effective neighborhood aggregation, while LM-centric approaches lack the topological context necessary to explicitly infer these hidden relationships. To test the robustness of our model against missing structural links, we simulate this exact lack of structural coherence by deleting valuable hub edges, which are edges connecting semantically coherent same-class nodes with high structural proximity where $\Pi_{ij} > 0.3$. We randomly remove $r \in (10\%, 20\%, 30\%)$ of these critical structural pathways from the adjacency matrix while perfectly preserving all node features and semantic texts. Results are shown in Tables \ref{tab:baselines_msl_10}, \ref{tab:baselines_msl_20}, and \ref{tab:baselines_msl_30}.

\begin{table}[htbp]
    \centering
    \caption{Performance comparison of different methods on Reddit, Instagram, and Wikics datasets under 10\% edge deletion.}
    \begin{tabular}{lcccccc}
        \toprule
        \multicolumn{1}{l}{\multirow{2}[4]{*}{Methods}} & \multicolumn{2}{c}{Reddit}          & \multicolumn{2}{c}{Instagram}       & \multicolumn{2}{c}{Wikics}                                                                                                                            \\
        \cmidrule(lr){2-3} \cmidrule(lr){4-5} \cmidrule(lr){6-7}
                                                        & Acc                                 & F1                                  & \multicolumn{1}{c}{Acc}             & \multicolumn{1}{c}{F1}              & \multicolumn{1}{c}{Acc}             & \multicolumn{1}{c}{F1}              \\
        \midrule
        GCN                                             & 49.58                               & 48.68                               & 59.27                               & 46.75                               & 70.38                               & 64.49                               \\
        GAT                                             & 50.06                               & 49.53                               & 57.01                               & 49.47                               & 69.46                               & 69.44                               \\
        \rowcolor[rgb]{ .851,  .851,  .851} Sent-BERT   & 55.99                               & 55.77                               & 61.04                               & 54.79                               & 74.20                               & 72.27                               \\
        \rowcolor[rgb]{ .851,  .851,  .851} RoBERTa     & 55.85                               & 55.78                               & 58.56                               & 53.11                               & 74.77                               & 74.49                               \\
        TAPE                                            & 57.16                               & 54.66                               & 64.61                               & 45.25                               & \textcolor{blue}{\underline{82.19}} & \textcolor{blue}{\underline{77.21}} \\
        ZeroG                                           & 59.17                               & 57.16                               & 66.02                               & 54.14                               & 76.79                               & 74.58                               \\
        LLaGA                                           & \textcolor{blue}{\underline{59.64}} & \textcolor{blue}{\underline{59.03}} & 61.21                               & 50.15                               & 68.91                               & 63.92                               \\
        InstructGLM                                     & 53.45                               & 49.97                               & 56.84                               & 48.98                               & 73.24                               & 68.99                               \\
        FLAG                                            & 54.65                               & 54.21                               & \textcolor{blue}{\underline{67.20}} & \textcolor{blue}{\underline{55.65}} & 78.45                               & 74.54                               \\
        Patton                                          & 56.06                               & 53.67                               & 60.93                               & 51.07                               & 77.98                               & 73.46                               \\
        CO-EVOLVE                                       & \textcolor{red}{\textbf{61.73}}     & \textcolor{red}{\textbf{60.51}}     & \textcolor{red}{\textbf{69.35}}     & \textcolor{red}{\textbf{56.64}}     & \textcolor{red}{\textbf{85.02}}     & \textcolor{red}{\textbf{81.39}}     \\
        \bottomrule
    \end{tabular}%
    \label{tab:baselines_msl_10}%
\end{table}%

Compared to the catastrophic degradation seen in GNN-centric models, CO-EVOLVE maintains robust performance across all three edge deletion rates. For instance, at a 30\% deletion rate on Reddit, GCN's accuracy plummets by 18.82\%, i.e., from 53.87\% to 35.05\%, while CO-EVOLVE only experiences a minimal 1.79\% drop, i.e., from 61.97\% to 60.18\%. Notably, while LM-only models, e.g., Sent-BERT and RoBERTa, exhibit zero degradation since they ignore the graph topology, their absolute performance remains significantly bounded below CO-EVOLVE's capabilities.

\begin{table}[htbp]
    \centering
    \caption{Performance comparison of different methods on Reddit, Instagram, and Wikics datasets under 20\% edge deletion.}
    \begin{tabular}{lcccccc}
        \toprule
        \multicolumn{1}{l}{\multirow{2}[4]{*}{Methods}} & \multicolumn{2}{c}{Reddit}          & \multicolumn{2}{c}{Instagram}       & \multicolumn{2}{c}{Wikics}                                                                                                                            \\
        \cmidrule(lr){2-3} \cmidrule(lr){4-5} \cmidrule(lr){6-7}
                                                        & Acc                                 & F1                                  & \multicolumn{1}{c}{Acc}             & \multicolumn{1}{c}{F1}              & \multicolumn{1}{c}{Acc}             & \multicolumn{1}{c}{F1}              \\
        \midrule
        GCN                                             & 42.42                               & 41.53                               & 51.37                               & 45.58                               & 61.28                               & 59.57                               \\
        GAT                                             & 44.57                               & 39.92                               & 49.85                               & 41.03                               & 63.42                               & 63.23                               \\
        \rowcolor[rgb]{ .851,  .851,  .851} Sent-BERT   & 55.99                               & 55.77                               & 61.04                               & 54.79                               & 74.20                               & 72.27                               \\
        \rowcolor[rgb]{ .851,  .851,  .851} RoBERTa     & 55.85                               & 55.78                               & 58.56                               & 53.11                               & 74.77                               & 74.49                               \\
        TAPE                                            & 51.36                               & 51.18                               & 60.95                               & 43.47                               & \textcolor{blue}{\underline{77.88}} & \textcolor{blue}{\underline{76.04}} \\
        ZeroG                                           & \textcolor{blue}{\underline{57.34}} & \textcolor{blue}{\underline{54.26}} & 60.60                               & 53.24                               & 68.54                               & 68.43                               \\
        LLaGA                                           & 55.95                               & 52.23                               & 57.26                               & 43.42                               & 62.97                               & 59.75                               \\
        InstructGLM                                     & 48.58                               & 44.08                               & 54.82                               & 48.21                               & 67.10                               & 65.34                               \\
        FLAG                                            & 50.86                               & 50.22                               & \textcolor{blue}{\underline{62.21}} & \textcolor{blue}{\underline{54.11}} & 74.36                               & 70.46                               \\
        Patton                                          & 54.10                               & 52.77                               & 60.22                               & 48.41                               & 73.50                               & 67.95                               \\
        CO-EVOLVE                                       & \textcolor{red}{\textbf{61.06}}     & \textcolor{red}{\textbf{59.81}}     & \textcolor{red}{\textbf{68.68}}     & \textcolor{red}{\textbf{55.95}}     & \textcolor{red}{\textbf{84.16}}     & \textcolor{red}{\textbf{80.48}}     \\
        \bottomrule
    \end{tabular}%
    \label{tab:baselines_msl_20}%
\end{table}%

CO-EVOLVE's resilience is driven by its Semantic Structure Learner and Dual-View gating mechanism. When critical structural hub edges are deleted, the message-passing pathways of standard GNNs are crippled. However, CO-EVOLVE leverages the pristine semantic text to identify high semantic similarity among these artificially disconnected nodes. The positive margin loss then actively reconstructs these missing links within the learned semantic graph. Simultaneously, the adaptive gate dynamically shifts the model's trust away from the corrupted, sparse static graph ($\mathbf{A}$) towards the newly imputed, semantically-rich graph ($\mathbf{S}$).

\begin{table}[htbp]
    \centering
    \caption{Performance comparison of different methods on Reddit, Instagram, and Wikics datasets under 30\% edge deletion.}
    \begin{tabular}{lcccccc}
        \toprule
        \multicolumn{1}{l}{\multirow{2}[4]{*}{Methods}} & \multicolumn{2}{c}{Reddit}          & \multicolumn{2}{c}{Instagram}       & \multicolumn{2}{c}{Wikics}                                                                                                                            \\
        \cmidrule(lr){2-3} \cmidrule(lr){4-5} \cmidrule(lr){6-7}
                                                        & Acc                                 & F1                                  & \multicolumn{1}{c}{Acc}             & \multicolumn{1}{c}{F1}              & \multicolumn{1}{c}{Acc}             & \multicolumn{1}{c}{F1}              \\
        \midrule
        GCN                                             & 35.05                               & 33.84                               & 41.89                               & 33.52                               & 50.80                               & 49.64                               \\
        GAT                                             & 36.15                               & 32.18                               & 44.09                               & 36.16                               & 49.67                               & 48.95                               \\
        \rowcolor[rgb]{ .851,  .851,  .851} Sent-BERT   & 55.99                               & 55.77                               & 61.04                               & 54.79                               & 74.20                               & 72.27                               \\
        \rowcolor[rgb]{ .851,  .851,  .851} RoBERTa     & 55.85                               & 55.78                               & 58.56                               & 53.11                               & 74.77                               & 74.49                               \\
        TAPE                                            & 51.01                               & 49.16                               & 55.06                               & 40.15                               & \textcolor{blue}{\underline{70.68}} & \textcolor{blue}{\underline{70.14}} \\
        ZeroG                                           & \textcolor{blue}{\underline{52.75}} & 45.96                               & \textcolor{blue}{\underline{57.91}} & \textcolor{blue}{\underline{48.64}} & 67.31                               & 62.71                               \\
        LLaGA                                           & 50.94                               & \textcolor{blue}{\underline{50.22}} & 52.17                               & 42.99                               & 53.93                               & 53.10                               \\
        InstructGLM                                     & 43.22                               & 39.88                               & 50.36                               & 42.70                               & 64.06                               & 57.98                               \\
        FLAG                                            & 49.98                               & 47.29                               & 53.90                               & 48.49                               & 67.78                               & 63.76                               \\
        Patton                                          & 48.61                               & 48.56                               & 55.84                               & 44.12                               & 67.84                               & 64.59                               \\
        CO-EVOLVE                                       & \textcolor{red}{\textbf{60.18}}     & \textcolor{red}{\textbf{58.49}}     & \textcolor{red}{\textbf{67.47}}     & \textcolor{red}{\textbf{54.85}}     & \textcolor{red}{\textbf{82.85}}     & \textcolor{red}{\textbf{78.93}}     \\
        \bottomrule
    \end{tabular}%
    \label{tab:baselines_msl_30}%
\end{table}%

By explicitly detecting semantic coherence and dynamically rewiring the graph topology, CO-EVOLVE successfully overcomes severe structural sparsity. This demonstrates the framework's superior ability to recover missing high-order structure via adaptive bidirectional fusion.

\subsection{Ablation Study (Q4)}
In addition to the micro experiments of key components conducted in Section \ref{sec:method}, we further evaluate the contribution of each core mechanism in CO-EVOLVE by formulating an ablation study isolating its four main components. Specifically, we evaluate the model w/o Soft Prompts (w/o SP), which removes the structural guidance condition from the LLM prompt; w/o Semantic Structure Learner (w/o SSL), which fixes the GNN topology to the original static graph instead of dynamically updating it with the learned semantic graph; w/o Conflict-Aware Loss (w/o CAL), which disables the mitigation of semantic-structural dissonance; and w/o Uncertainty-Gated Consistency (w/o UGC), which removes the entropy-weighted bidirectional consistency loss. By comparing these isolated variants, we intend to confirm that bidirectional error correction and adaptive fusion are collectively crucial in overcoming semantic-structural dissonance learning. Results are shown in Table \ref{tab:ablation_study}.

\begin{table}[htbp]
    \centering
    \caption{Ablation studies on key components of CO-EVOLVE on Reddit, Instagram, and Wikics datasets.}
    \begin{tabular}{lcccccc}
        \toprule
        \multicolumn{1}{l}{\multirow{2}[4]{*}{Methods}} & \multicolumn{2}{c}{Reddit}      & \multicolumn{2}{c}{Instagram}   & \multicolumn{2}{c}{Wikics}                                                                                                            \\
        \cmidrule(lr){2-3} \cmidrule(lr){4-5} \cmidrule(lr){6-7}
                                                        & Acc                             & F1                              & \multicolumn{1}{c}{Acc}         & \multicolumn{1}{c}{F1}          & \multicolumn{1}{c}{Acc}         & \multicolumn{1}{c}{F1}          \\
        \midrule
        w/o SP                                          & 60.13                           & 58.61                           & 67.21                           & 55.42                           & 84.18                           & 79.52                           \\
        w/o SSL                                         & 56.46                           & 55.25                           & 64.55                           & 54.25                           & 80.29                           & 76.86                           \\
        w/o CAL                                         & 57.85                           & 56.32                           & 65.84                           & 52.89                           & 81.64                           & 78.17                           \\
        w/o UGC                                         & 59.42                           & 59.17                           & 68.32                           & 56.05                           & 83.11                           & 80.65                           \\
        CO-EVOLVE                                       & \textcolor{red}{\textbf{61.97}} & \textcolor{red}{\textbf{60.84}} & \textcolor{red}{\textbf{69.74}} & \textcolor{red}{\textbf{57.11}} & \textcolor{red}{\textbf{85.35}} & \textcolor{red}{\textbf{81.89}} \\
        \bottomrule
    \end{tabular}%
    \label{tab:ablation_study}%
\end{table}%

We observe a consistent and significant performance degradation across all three datasets when any of the four core mechanisms are isolated. The most severe drops in both Accuracy and F1 score occur when fixing the GNN topology (w/o SSL) and removing the mitigation of semantic-structural dissonance (w/o CAL), followed by noticeable decreases without structural guidance (w/o SP) and uncertainty-gated consistency (w/o UGC). This uniformly indicates that no single component is solely responsible for CO-EVOLVE's performance. Rather, they serve interdependent roles.

Specifically, removing the Semantic Structure Learner causes the most dramatic drop, a over $5\%$ absolute reduction in F1 on Wikics and Reddit, proving that relying purely on the static, heterophilous graph traps the GNN in noisy topologies and prevents the recovery of hidden homophilous links. Disabling the Conflict-Aware Loss results in the second-largest degradation, validating that without explicit penalties, false semantic friends successfully merge in the embedding space, causing bidirectional error propagation. Removing the structural guidance condition leads to an approximate $2\sim3\%$ performance reduction, confirming that the LLM suffers from spatial blindness and hallucinates semantic clusters that ignore true topological boundaries without soft prompt grounding. Finally, removing the Uncertainty-Gated Consistency yields a $1\sim2\%$ drop; without entropy-weighted verification, noisy signals are blindly accepted during the co-evolutionary alternating optimization phase, degrading the stability of mutual enhancement.

These results confirm that bidirectional error correction and adaptive fusion are collectively superior and deeply coupled. Without the synergistic interaction of structurally-grounded semantics, semantically-corrected topology, conflict resolution, and uncertainty gating, the framework loses its resilience against semantic-structural dissonance.

\subsection{Hyper-parameter Sensitivity (Q5)}
Finally, we analyze the sensitivity of the CO-EVOLVE framework's performance with respect to hyper-parameters. First, we vary the neighborhood pruning degree parameter $k$ used in the Semantic Structure Learner to assess the trade-off between adding valuable missing links versus introducing noise. Second, we evaluate CO-EVOLVE under varying values of the structural positive threshold $\alpha$ and the structural irrelevance threshold $\epsilon$ to understand their exact impact on the stability of the conflict-aware loss mechanism.

\begin{figure}[htbp]
    \centering
    \begin{subfigure}[t]{0.48\linewidth}
        \centering
        \includegraphics[width=\linewidth]{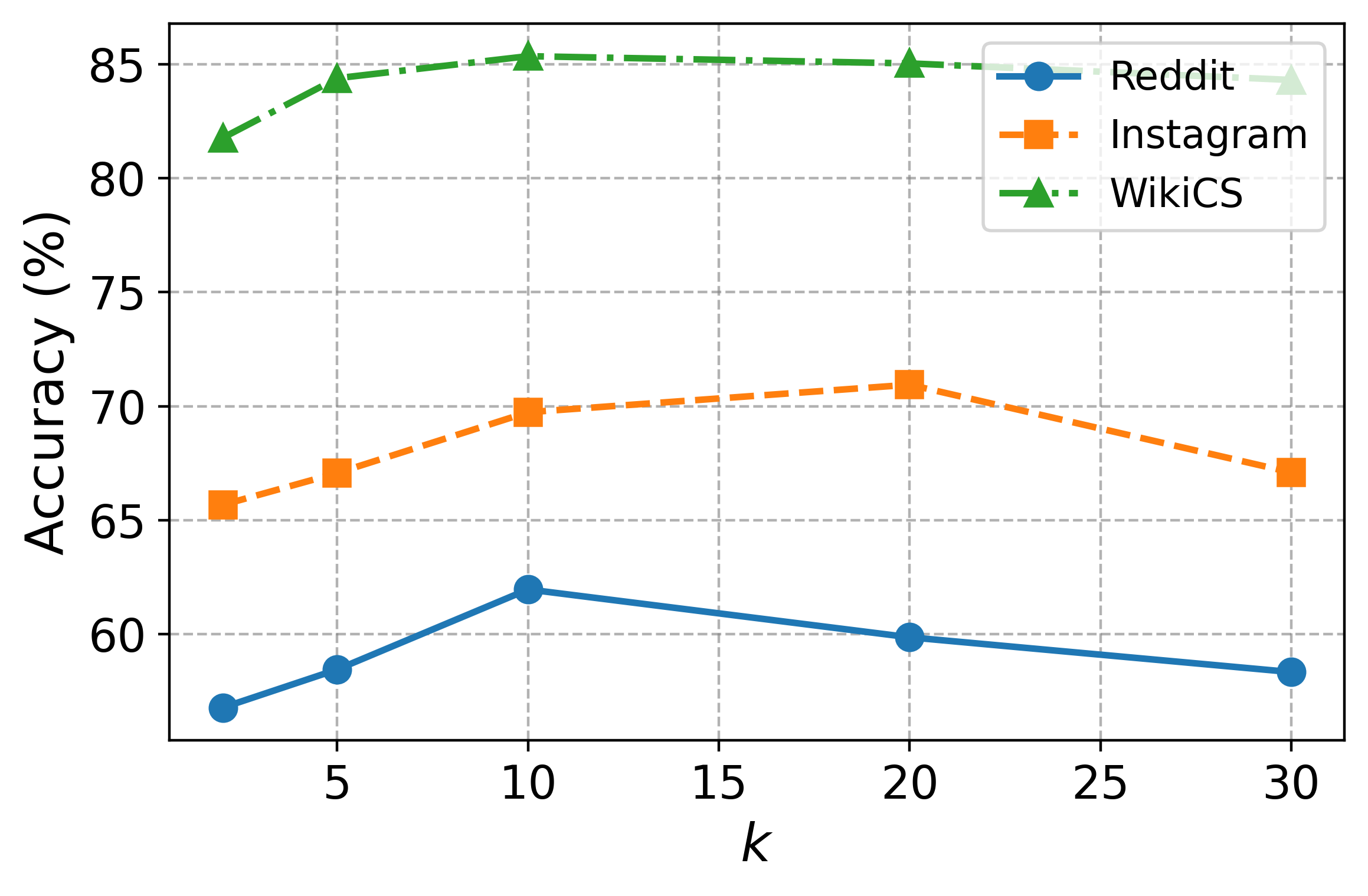}
        \caption{Accuracy w.r.t $k$}
    \end{subfigure}\hfill
    \begin{subfigure}[t]{0.48\linewidth}
        \centering
        \includegraphics[width=\linewidth]{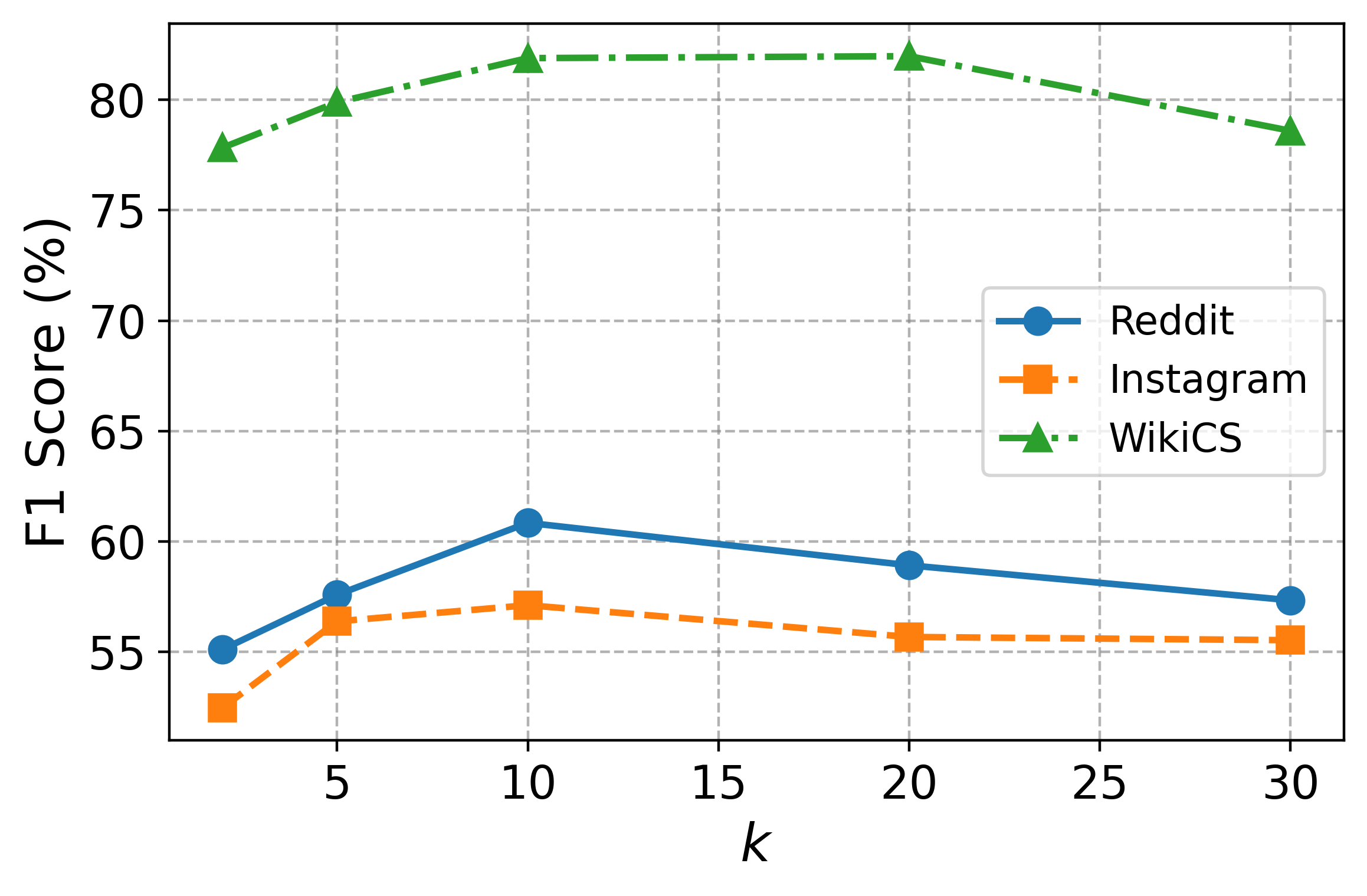}
        \caption{F1 score w.r.t $k$}
    \end{subfigure}
    \caption{Hyper-parameter sensitivity analysis of CO-EVOLVE on varying neighborhood size $k$.}
    \label{fig:hyper_param_k}
\end{figure}

\begin{figure}[htbp]
    \centering
    \begin{subfigure}[t]{0.48\linewidth}
        \centering
        \includegraphics[width=\linewidth]{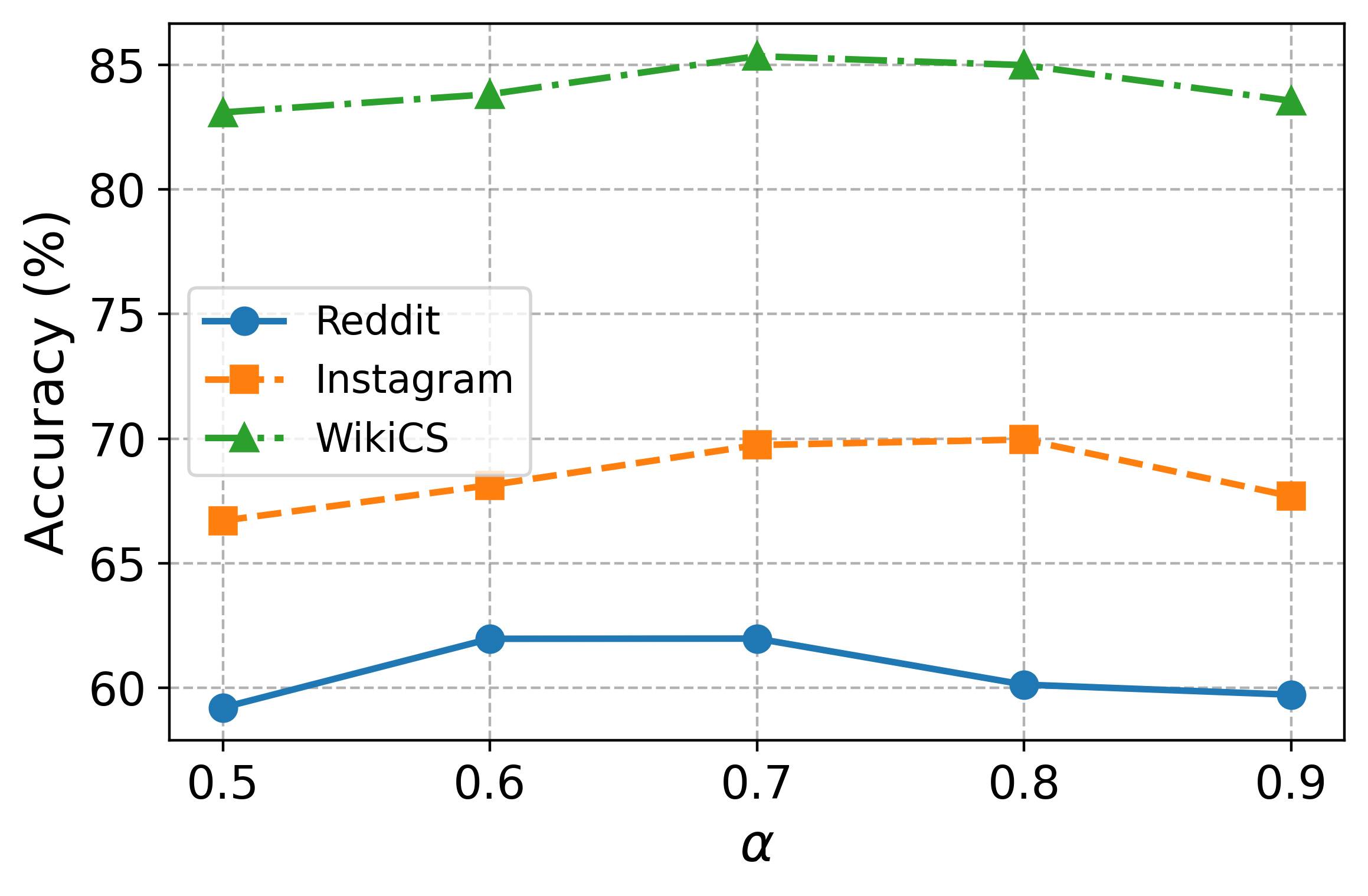}
        \caption{Accuracy w.r.t $\alpha$}
    \end{subfigure}\hfill
    \begin{subfigure}[t]{0.48\linewidth}
        \centering
        \includegraphics[width=\linewidth]{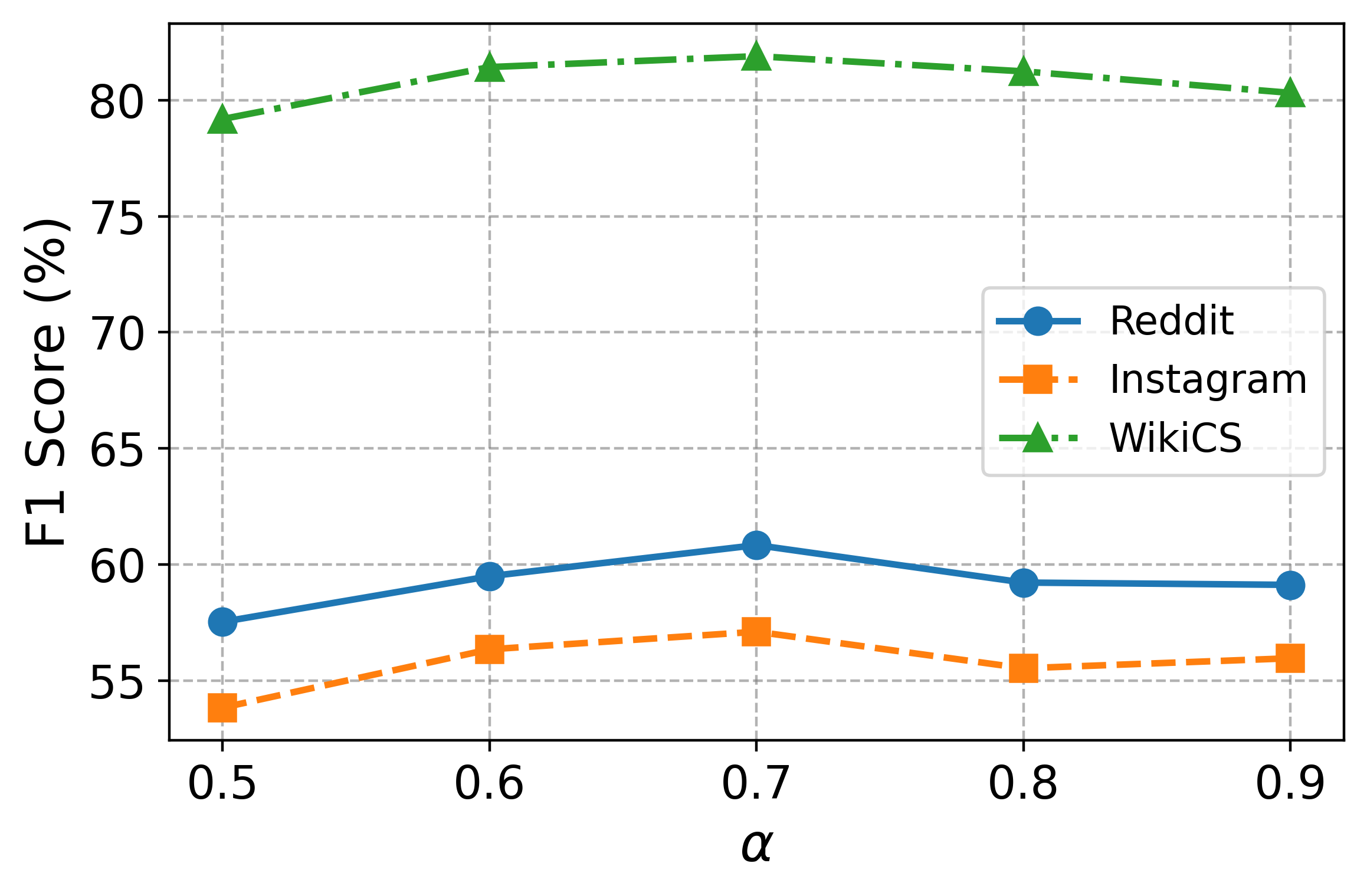}
        \caption{F1 score w.r.t $\alpha$}
    \end{subfigure}
    \caption{Hyper-parameter sensitivity analysis of CO-EVOLVE on varying structural positive threshold $\alpha$.}
    \label{fig:hyper_param_alpha}
\end{figure}

\begin{figure}[htbp]
    \centering
    \begin{subfigure}[t]{0.48\linewidth}
        \centering
        \includegraphics[width=\linewidth]{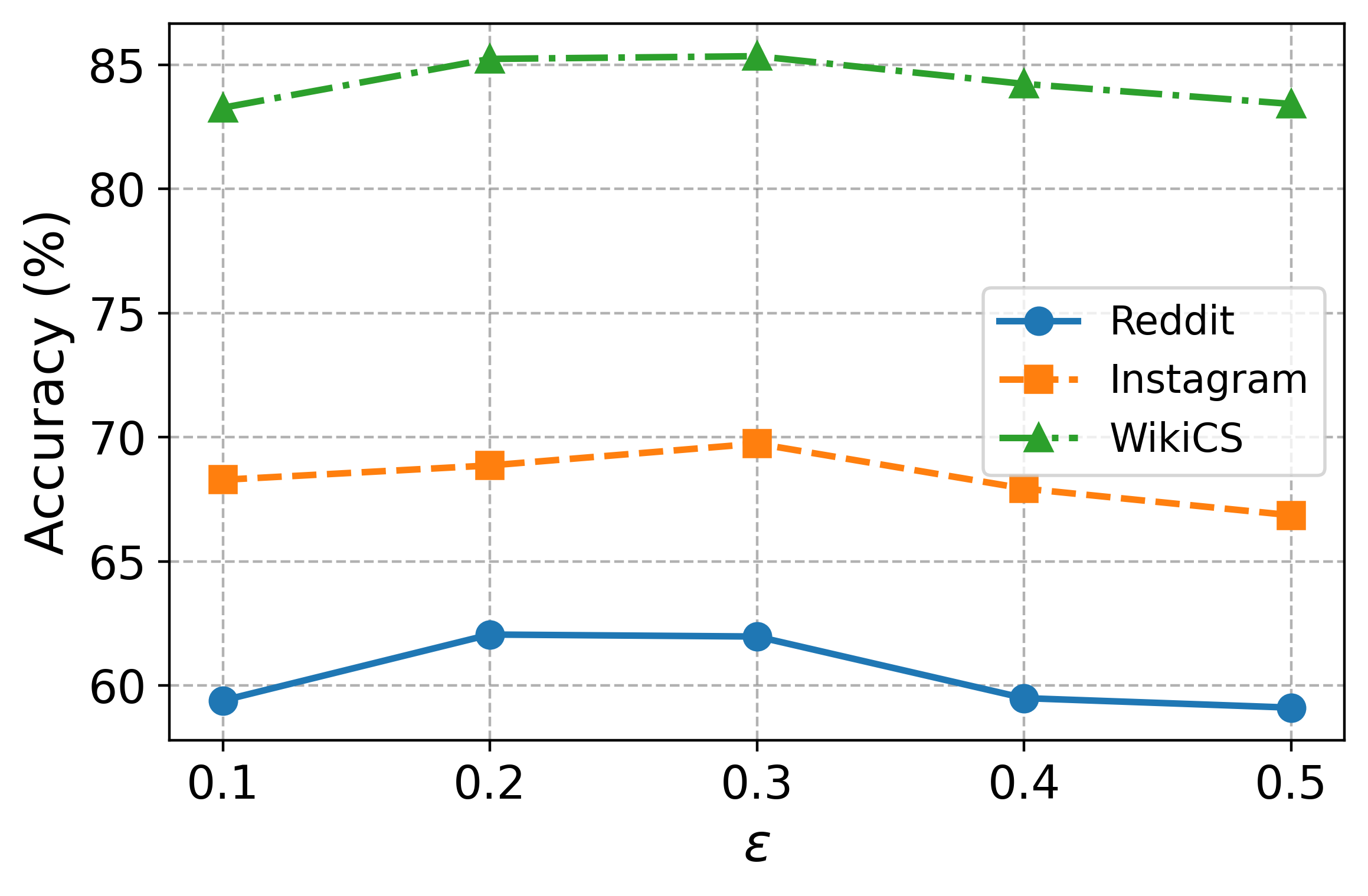}
        \caption{Accuracy w.r.t $\epsilon$}
    \end{subfigure}\hfill
    \begin{subfigure}[t]{0.48\linewidth}
        \centering
        \includegraphics[width=\linewidth]{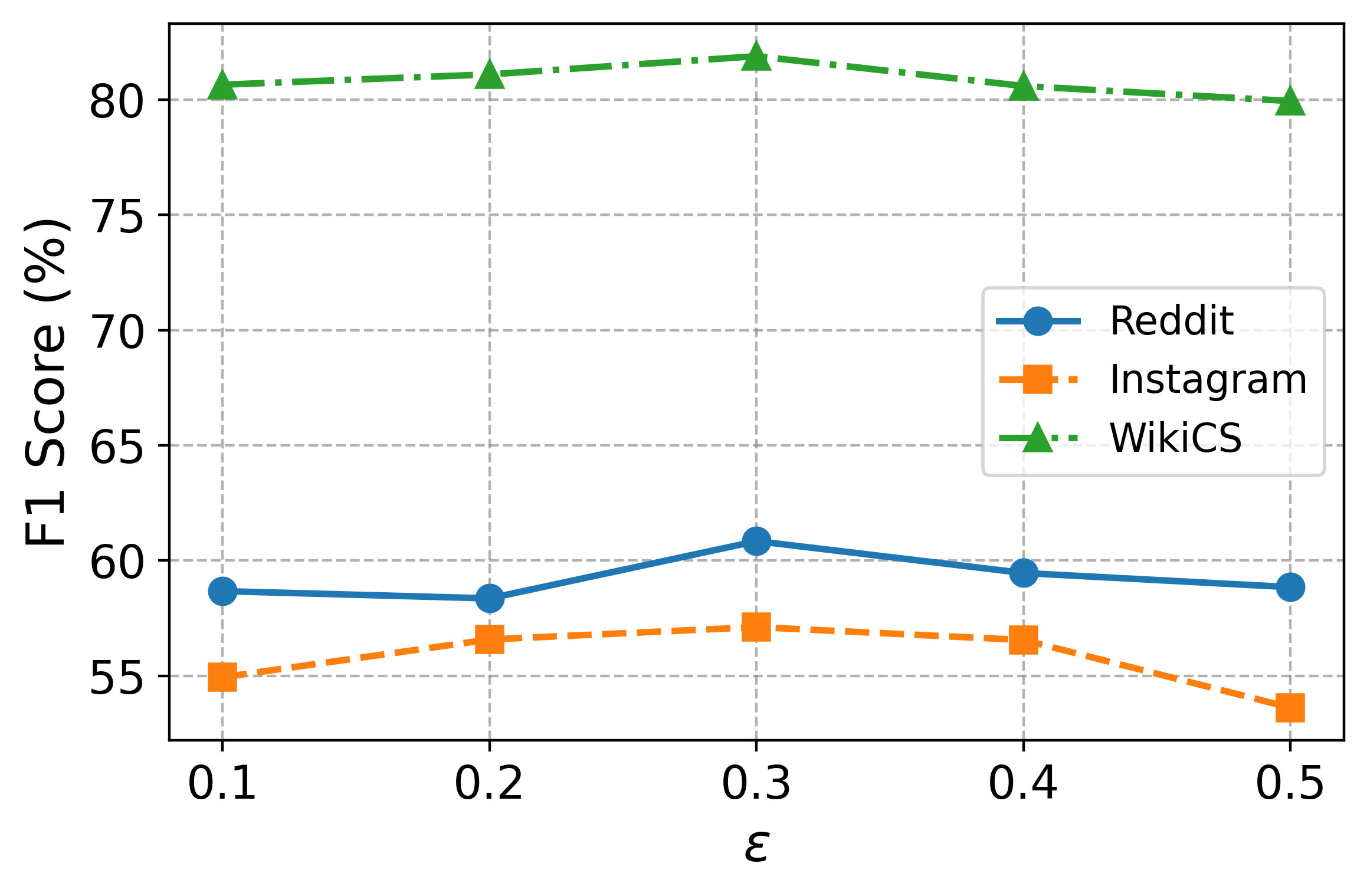}
        \caption{F1 score w.r.t $\epsilon$}
    \end{subfigure}
    \caption{Hyper-parameter sensitivity analysis of CO-EVOLVE on varying structural irrelevance threshold $\epsilon$.}
    \label{fig:hyper_param_epsilon}
\end{figure}

We observe a inverted-U-like performance trend across the three datasets when varying the neighborhood size $k$, the structural positive threshold $\alpha$, and the structural irrelevance threshold $\epsilon$, as shown in Figures~\ref{fig:hyper_param_k}, \ref{fig:hyper_param_alpha}, and \ref{fig:hyper_param_epsilon}. Rather than strictly achieving peak performance at our default settings ($k=10$, $\alpha=0.7$, $\epsilon=0.3$) in every individual case, we select this configuration because it consistently yields the best overall performance and stability across structurally diverse scenarios. Nonetheless, the framework exhibits graceful degradation at extreme values, demonstrating that moderately tuned parameters effectively balance structural reconstruction and noise filtration.

Specifically, regarding the semantic neighborhood pruning degree $k$, we observe that performance drops sharply at small values and degrades more gradually at large values. This occurs because a small $k$ creates an overly sparse graph that fails to fully recover valuable missing homophilous links, whereas a large $k$ introduces weak semantic correlations and topological noise that limit message-passing efficiency. Similarly, for the structural positive threshold $\alpha$, performance diminishes when moving away from the optimal 0.6-0.8 range. Setting it too low, i.e., $\alpha=0.5$ erroneously pulls noisy neighbors into the positive set, while overly strict configurations provide insufficient true positive supervision to adequately warp the semantic manifold. Finally, for the structural irrelevance threshold $\epsilon$ responsible for penalizing false semantic friends, performance dips at strict cutoffs and relaxes too much at high values. A strict cutoff overlooks deceptive cross-class pairs, while a relaxed threshold falsely penalizes valid, moderately distant structural neighbors. Therefore, calibrated thresholds are essential to prevent semantic-structural dissonance without triggering false conflict signals.

CO-EVOLVE relies on optimal structural density and accurate conflict mining to achieve peak performance, but it maintains highly competitive baseline performance across a broad spectrum of parameter variations, underscoring its inherent robustness.

\section{Conclusion}
In this paper, we challenge the conventional static, unidirectional paradigms used to integrate LLMs and GNNs, demonstrating that they inherently suffer from bidirectional error propagation, semantic-structural dissonance, and blind alignment. To overcome these critical bottlenecks, we propose CO-EVOLVE, a dual-view co-evolution framework that treats graph topology and semantic embeddings as dynamic, mutually reinforcing latent variables. By establishing a cyclic feedback loop via alternating optimization, CO-EVOLVE enables unprecedented mutual correction between the modalities: the GNN injects topological context as soft prompts to guide the LLM, while the LLM simultaneously constructs an adaptive dynamic semantic graph to refine the GNN's topological view. Together with a Hard-Structure Conflict-Aware Contrastive Loss and an Uncertainty-Gated Consistency strategy, our framework fundamentally resolves the dissonance between semantic similarity and structural connectivity, preserving robustness in heterophilous and noisy environments. Extensive empirical evaluations on public benchmarks validate that CO-EVOLVE achieves state-of-the-art performance. Crucially, rigorous stress tests reveal that CO-EVOLVE exhibits remarkable resilience against both deceptive semantic features, i.e., false semantic friends, and severe spatial sparsity, i.e., missing structural links. Ultimately, replacing rigid pipelines with mutually evolving latent variables provides a principled foundation for unified graph-language modeling, paving the way for more rigorous and reliable reasoning systems in complex, real-world networks.

\bibliographystyle{ieeetr}
\bibliography{main}

@article{chien2021node,
  title   = {Node feature extraction by self-supervised multi-scale neighborhood prediction},
  author  = {Chien, Eli and Chang, Wei-Cheng and Hsieh, Cho-Jui and Yu, Hsiang-Fu and Zhang, Jiong and Milenkovic, Olgica and Dhillon, Inderjit S},
  journal = {arXiv preprint arXiv:2111.00064},
  year    = {2021}
}

@article{shehzad2026graph,
  title     = {Graph transformers: A survey},
  author    = {Shehzad, Ahsan and Xia, Feng and Abid, Shagufta and Peng, Ciyuan and Yu, Shuo and Zhang, Dongyu and Verspoor, Karin},
  journal   = {IEEE Transactions on Neural Networks and Learning Systems},
  year      = {2026},
  publisher = {IEEE}
}

@article{wang2025graph,
  title     = {Graph machine learning in the era of large language models (llms)},
  author    = {Wang, Shijie and Huang, Jiani and Chen, Zhikai and Song, Yu and Tang, Wenzhuo and Mao, Haitao and Fan, Wenqi and Liu, Hui and Liu, Xiaorui and Yin, Dawei and others},
  journal   = {ACM Transactions on Intelligent Systems and Technology},
  volume    = {16},
  number    = {5},
  pages     = {1--40},
  year      = {2025},
  publisher = {ACM New York, NY}
}

@article{wang2025comprehensive,
  title   = {A comprehensive survey in llm (-agent) full stack safety: Data, training and deployment},
  author  = {Wang, Kun and Zhang, Guibin and Zhou, Zhenhong and Wu, Jiahao and Yu, Miao and Zhao, Shiqian and Yin, Chenlong and Fu, Jinhu and Yan, Yibo and Luo, Hanjun and others},
  journal = {arXiv preprint arXiv:2504.15585},
  year    = {2025}
}

@inproceedings{yang2025flag,
  title     = {Flag: Fraud detection with llm-enhanced graph neural network},
  author    = {Yang, Chengdong and Liu, Hongrui and Wang, Daixin and Zhang, Zhiqiang and Yang, Cheng and Shi, Chuan},
  booktitle = {Proceedings of the 31st ACM SIGKDD Conference on Knowledge Discovery and Data Mining V. 2},
  pages     = {5150--5160},
  year      = {2025}
}

@article{chen2025beyond,
  title     = {Beyond topology-based graph mining: Deep analysis research networks via evolutionary topology and content fusion},
  author    = {Chen, Xueyu and Miao, Ran and Hu, Liang and Zhang, Qi and Song, Kaitao and Naseem, Usman and Zhao, Cairong},
  journal   = {Information Fusion},
  pages     = {103922},
  year      = {2025},
  publisher = {Elsevier}
}

@inproceedings{wang2025bridging,
  title     = {Bridging Molecular Graphs and Large Language Models},
  author    = {Wang, Runze and Yang, Mingqi and Shen, Yanming},
  booktitle = {Proceedings of the AAAI Conference on Artificial Intelligence},
  volume    = {39},
  pages     = {21234--21242},
  year      = {2025}
}

@article{zhu2025llm,
  title   = {Llm as gnn: Graph vocabulary learning for text-attributed graph foundation models},
  author  = {Zhu, Xi and Xue, Haochen and Zhao, Ziwei and Xu, Wujiang and Huang, Jingyuan and Guo, Minghao and Wang, Qifan and Zhou, Kaixiong and Razzak, Imran and Zhang, Yongfeng},
  journal = {arXiv preprint arXiv:2503.03313},
  year    = {2025}
}

@article{ju2025survey,
  title     = {A survey of graph neural networks in real world: Imbalance, noise, privacy and ood challenges},
  author    = {Ju, Wei and Yi, Siyu and Wang, Yifan and Xiao, Zhiping and Mao, Zhengyang and Li, Hourun and Gu, Yiyang and Qin, Yifang and Yin, Nan and Wang, Senzhang and others},
  journal   = {IEEE Transactions on Pattern Analysis and Machine Intelligence},
  year      = {2025},
  publisher = {IEEE}
}

@inproceedings{kipf2017semi,
  title     = {Semi-Supervised Classification with Graph Convolutional Networks},
  author    = {Kipf, Thomas N and Welling, Max},
  booktitle = {International Conference on Learning Representations},
  year      = {2017}
}

@article{velickovic2018graph,
  title   = {Graph Attention Networks},
  author  = {Veli{\v{c}}kovi{\'c}, Petar and Cucurull, Guillem and Casanova, Arantxa and Romero, Adriana and Lio, Pietro and Bengio, Yoshua},
  journal = {arXiv preprint arXiv:1710.10903},
  year    = {2017}
}

@inproceedings{vaswani2017attention,
  title     = {Attention is all you need},
  author    = {Vaswani, Ashish and Shazeer, Noam and Parmar, Niki and Uszkoreit, Jakob and Jones, Llion and Gomez, Aidan N and Kaiser, {\L}ukasz and Polosukhin, Illia},
  booktitle = {Advances in neural information processing systems},
  pages     = {5998--6008},
  year      = {2017}
}

@inproceedings{brown2020language,
  title     = {Language models are few-shot learners},
  author    = {Brown, Tom and Mann, Benjamin and Ryder, Nick and Subbiah, Melanie and Kaplan, Jared D and Dhariwal, Prafulla and Neelakantan, Arvind and Shyam, Pranav and Sastry, Girish and Askell, Amanda and others},
  booktitle = {Advances in neural information processing systems},
  volume    = {33},
  pages     = {1877--1901},
  year      = {2020}
}

@article{liu2024deepseek,
  title   = {Deepseek-v3 technical report},
  author  = {Liu, Aixin and Feng, Bei and Xue, Bing and Wang, Bingxuan and Wu, Bochao and Lu, Chengda and Zhao, Chenggang and Deng, Chengqi and Zhang, Chenyu and Ruan, Chong and others},
  journal = {arXiv preprint arXiv:2412.19437},
  year    = {2024}
}

@article{ren2024survey,
  title   = {A Survey of Large Language Models for Graphs},
  author  = {Ren, Xubin and Tang, Jiabin and Yin, Dawei and Chawla, Nitesh and Huang, Chao},
  journal = {arXiv preprint arXiv:2405.08011},
  year    = {2024}
}

@inproceedings{he2024harnessing,
  title     = {Harnessing Explanations: LLM-to-LM Interpreter for Enhanced Text-Attributed Graph Representation Learning},
  author    = {He, Xiaoxin and others},
  booktitle = {International Conference on Learning Representations},
  year      = {2024}
}

@article{liu2019roberta,
  title   = {Roberta: A robustly optimized bert pretraining approach},
  author  = {Liu, Yinhan and Ott, Myle and Goyal, Naman and Du, Jingfei and Joshi, Mandar and Chen, Danqi and Levy, Omer and Lewis, Mike and Zettlemoyer, Luke and Stoyanov, Veselin},
  journal = {arXiv preprint arXiv:1907.11692},
  year    = {2019}
}

@article{he2023harnessing,
  title   = {Harnessing explanations: Llm-to-lm interpreter for enhanced text-attributed graph representation learning},
  author  = {He, Xiaoxin and Bresson, Xavier and Laurent, Thomas and Perold, Adam and LeCun, Yann and Hooi, Bryan},
  journal = {arXiv preprint arXiv:2305.19523},
  year    = {2023}
}

@article{grattafiori2024llama,
  title   = {The llama 3 herd of models},
  author  = {Grattafiori, Aaron and Dubey, Abhimanyu and Jauhri, Abhinav and Pandey, Abhinav and Kadian, Abhishek and Al-Dahle, Ahmad and Letman, Aiesha and Mathur, Akhil and Schelten, Alan and Vaughan, Alex and others},
  journal = {arXiv preprint arXiv:2407.21783},
  year    = {2024}
}

@article{hu2022lora,
  title   = {Lora: Low-rank adaptation of large language models.},
  author  = {Hu, Edward J and Shen, Yelong and Wallis, Phillip and Allen-Zhu, Zeyuan and Li, Yuanzhi and Wang, Shean and Wang, Liang and Chen, Weizhu and others},
  journal = {Iclr},
  volume  = {1},
  number  = {2},
  pages   = {3},
  year    = {2022}
}

@inproceedings{reimers2019sentence,
  title     = {Sentence-bert: Sentence embeddings using siamese bert-networks},
  author    = {Reimers, Nils and Gurevych, Iryna},
  booktitle = {Proceedings of the 2019 conference on empirical methods in natural language processing and the 9th international joint conference on natural language processing (EMNLP-IJCNLP)},
  pages     = {3982--3992},
  year      = {2019}
}

@article{liu2023one,
  title   = {One for All: Towards Training One Graph Model for All Classification Tasks},
  author  = {Liu, Hao and Feng, Jiarui and Kong, Lecheng and Liang, Ningyue and Tao, Dacheng and Chen, Yixin and Zhang, Muhan},
  journal = {arXiv preprint arXiv:2310.00149},
  year    = {2023}
}

@inproceedings{wei2024llmrec,
  title     = {LLMRec: Large Language Models with Graph Augmentation for Recommendation},
  author    = {Wei, Wei and others},
  booktitle = {Proceedings of the ACM Web Conference 2024},
  year      = {2024}
}

@inproceedings{yu2025samgpt,
  title     = {Samgpt: Text-free graph foundation model for multi-domain pre-training and cross-domain adaptation},
  author    = {Yu, Xingtong and Gong, Zechuan and Zhou, Chang and Fang, Yuan and Zhang, Hui},
  booktitle = {Proceedings of the ACM on Web Conference 2025},
  pages     = {1142--1153},
  year      = {2025}
}

@inproceedings{zheng2025enhancing,
  title     = {Enhancing Homophily-Heterophily Separation: Relation-Aware Learning in Heterogeneous Graphs},
  author    = {Zheng, Ziyu and Yang, Yaming and Guan, Ziyu and Zhao, Wei and Lu, Weigang},
  booktitle = {Proceedings of the 31st ACM SIGKDD Conference on Knowledge Discovery and Data Mining V. 2},
  pages     = {4050--4061},
  year      = {2025}
}

@inproceedings{jin2020graph,
  title     = {Graph structure learning for robust graph neural networks},
  author    = {Jin, Wei and Ma, Yao and Liu, Xiaorui and Tang, Xianfeng and Wang, Suhang and Tang, Jiliang},
  booktitle = {Proceedings of the 26th ACM SIGKDD international conference on knowledge discovery \& data mining},
  pages     = {66--74},
  year      = {2020}
}

@article{li2024glbench,
  title   = {Glbench: A comprehensive benchmark for graph with large language models},
  author  = {Li, Yuhan and Wang, Peisong and Zhu, Xiao and Chen, Aochuan and Jiang, Haiyun and Cai, Deng and Chan, Victor W and Li, Jia},
  journal = {Advances in Neural Information Processing Systems},
  volume  = {37},
  pages   = {42349--42368},
  year    = {2024}
}

@article{fang2024gaugl,
  title   = {GAugLLM: Improving Graph Contrastive Learning for Text-Attributed Graphs with Large Language Models},
  author  = {Fang, Yi and Li, Xuming and Li, Xiaochun and Li, Bo},
  journal = {arXiv preprint arXiv:2406.11945},
  year    = {2024}
}

@article{jin2023edgeformers,
  title   = {Edgeformers: Graph-Empowered Transformers for Representation Learning on Textual-Edge Networks},
  author  = {Jin, Bowen and others},
  journal = {arXiv preprint arXiv:2302.11042},
  year    = {2023}
}

@article{chen2024llaga,
  title   = {LLaGA: Large Language and Graph Assistant},
  author  = {Chen, Y and others},
  journal = {arXiv preprint arXiv:2402.08170},
  year    = {2024}
}

@inproceedings{huang2024can,
  title     = {Can gnn be good adapter for llms?},
  author    = {Huang, Xuanwen and Han, Kaiqiao and Yang, Yang and Bao, Dezheng and Tao, Quanjin and Chai, Ziwei and Zhu, Qi},
  booktitle = {Proceedings of the ACM Web Conference 2024},
  pages     = {893--904},
  year      = {2024}
}

@inproceedings{ye2024language,
  title     = {Language is all a graph needs},
  author    = {Ye, Ruosong and Zhang, Caiqi and Wang, Runhui and Xu, Shuyuan and Zhang, Yongfeng},
  booktitle = {Findings of the association for computational linguistics: EACL 2024},
  pages     = {1955--1973},
  year      = {2024}
}

@inproceedings{li2024zerog,
  title     = {Zerog: Investigating cross-dataset zero-shot transferability in graphs},
  author    = {Li, Yuhan and Wang, Peisong and Li, Zhixun and Yu, Jeffrey Xu and Li, Jia},
  booktitle = {Proceedings of the 30th ACM SIGKDD Conference on Knowledge Discovery and Data Mining},
  pages     = {1725--1735},
  year      = {2024}
}

@article{tang2023graphgpt,
  title   = {GraphGPT: Graph Instruction Tuning for Large Language Models},
  author  = {Tang, Jiabin and others},
  journal = {arXiv preprint arXiv:2310.13023},
  year    = {2023}
}

@article{jin2023patton,
  title   = {Patton: Language Model Pretraining on Text-Rich Networks},
  author  = {Jin, Bowen and others},
  journal = {arXiv preprint arXiv:2305.12268},
  year    = {2023}
}

@article{zhang2021scr,
  title   = {SCR: Training graph neural networks with consistency regularization},
  author  = {Zhang, Chenhui and He, Yufei and Cen, Yukuo and Hou, Zhenyu and Feng, Wenzheng and Dong, Yuxiao and Cheng, Xu and Cai, Hongyun and He, Feng and Tang, Jie},
  journal = {arXiv preprint arXiv:2112.04319},
  year    = {2021}
}

@article{fang2026knowledge,
  title     = {Knowledge distillation and dataset distillation of large language models: Emerging trends, challenges, and future directions},
  author    = {Fang, Luyang and Yu, Xiaowei and Cai, Jiazhang and Chen, Yongkai and Wu, Shushan and Liu, Zhengliang and Yang, Zhenyuan and Lu, Haoran and Gong, Xilin and Liu, Yufang and others},
  journal   = {Artificial Intelligence Review},
  volume    = {59},
  number    = {1},
  pages     = {17},
  year      = {2026},
  publisher = {Springer}
}

@inproceedings{zhu2020beyond,
  title     = {Beyond Homophily in Graph Neural Networks: Current Limitations and Effective Designs},
  author    = {Zhu, Jiong and Yan, Yujun and Zhao, Lingxiao and Heimann, Mark and Akoglu, Leman and Koutra, Danai},
  booktitle = {Advances in Neural Information Processing Systems},
  volume    = {33},
  pages     = {7793--7804},
  year      = {2020}
}

@inproceedings{bo2021beyond,
  title     = {Beyond Low-frequency Information in Graph Convolutional Networks},
  author    = {Bo, Deyu and Wang, Xiao and Shi, Chuan and Shen, Huawei},
  booktitle = {Proceedings of the AAAI Conference on Artificial Intelligence},
  volume    = {35},
  pages     = {3950--3957},
  year      = {2021}
}

@inproceedings{chien2021adaptive,
  title     = {Adaptive Universal Generalized PageRank Graph Neural Network},
  author    = {Chien, Eli and Peng, Jianhao and Li, Pan and Milenkovic, Olgica},
  booktitle = {International Conference on Learning Representations},
  year      = {2021}
}

@article{lim2021large,
  title   = {Large scale learning on non-homophilous graphs: New benchmarks and strong simple methods},
  author  = {Lim, Derek and Hohne, Felix and Li, Xiuyu and Huang, Sijia Linda and Gupta, Vaishnavi and Bhalerao, Omkar and Lim, Ser Nam},
  journal = {Advances in neural information processing systems},
  volume  = {34},
  pages   = {20887--20902},
  year    = {2021}
}

@inproceedings{wang2024understanding,
  title     = {Understanding Heterophily for Graph Neural Networks},
  author    = {Wang, Junfu and Guo, Yuanfang and Yang, Liang and Wang, Yunhong},
  booktitle = {International Conference on Machine Learning},
  year      = {2024}
}

@inproceedings{yang2024hlcl,
  title     = {Graph Contrastive Learning under Heterophily via Graph Filters},
  author    = {Yang, Wenhan and Mirzasoleiman, Baharan},
  booktitle = {Conference on Uncertainty in Artificial Intelligence},
  year      = {2024}
}

@inproceedings{wang2024hetergcl,
  title     = {HeterGCL: Graph Contrastive Learning Framework on Heterophilic Graph},
  author    = {Wang, Chenhao and Liu, Yong and Yang, Yan and Li, Wei},
  booktitle = {Proceedings of the Thirty-Third International Joint Conference on Artificial Intelligence},
  pages     = {2397--2405},
  year      = {2024}
}

@article{gong2026survey,
  title     = {A Survey on Learning from Graphs with Heterophily: Recent Advances and Future Directions},
  author    = {Gong, Chenghua and Cheng, Yao and Yu, Jianxiang and Xu, Can and Shan, Caihua and Luo, Siqiang and Li, Xiang},
  journal   = {Frontiers of Computer Science},
  volume    = {20},
  number    = {2},
  pages     = {2002314},
  year      = {2026},
  publisher = {Springer}
}

@article{zheng2022graph,
  title   = {Graph Neural Networks for Graphs with Heterophily: A Survey},
  author  = {Zheng, Xin and Wang, Yi and Liu, Yixin and Li, Ming and Zhang, Miao and Jin, Di and Yu, Philip S. and Pan, Shirui},
  journal = {arXiv preprint arXiv:2202.07082},
  year    = {2022}
}

\appendix
\subsection{Datasets Statistics}
\label{tab:appendix_dataset_statistics}
\begin{table}[htbp]
    \centering
    % \caption{Dataset statistics}
    \begin{tabular}{lrrrrr}
        \toprule
        Dataset   & \# Nodes & \# Edges & Avg. \# Tok & \# Classes & \# Train \\
        \midrule
        Reddit    & 33,434   & 198,448  & 203.84      & 2          & 10.00\%  \\
        Instagram & 11,339   & 144,010  & 59.25       & 2          & 10.00\%  \\
        WikiCS    & 11,701   & 216,123  & 642.04      & 10         & 4.96\%   \\
        \bottomrule
    \end{tabular}
\end{table}

\subsection{Pseudocode}
\begin{algorithm}[htbp]
    \caption{Dual-View Co-Evolution: Resolving Semantic-Structural Dissonance}
    \label{alg:framework}

    \KwIn{
        Graph $\mathcal{G} = (\mathcal{V}, {A})$; Raw Text $\mathcal{T}$;
        Node Features $\mathcal{X}$; Labels $\mathcal{Y}_L$; PPR Matrix ${\Pi}$ (pre-computed); Max Epochs $E$; Warm-up Epochs $E_{\text{warm}}$
    }
    \KwOut{Optimized parameters $\theta_{\text{LLM}}$, $\theta_{\text{GNN}}$}

    \BlankLine
    \textbf{Initialization:} Pre-train GNN on $({A}, {X})$ and LLM on $\mathcal{T}$ for $E_{\text{warm}}$ epochs. Initialize $\tilde{{A}} \leftarrow {A}$.

    \BlankLine
    \For{$e \leftarrow 1$ \KwTo $E$}{

    \tcc{Phase 1: Structural View (GNN)}
    ${H}_{\text{struct}}, P_{\text{GNN}}
        \leftarrow \text{GNN}_{\theta}(\tilde{{A}}, {X})$\;

    ${P}_{\text{struct}}
        \leftarrow \text{MLP}_{\text{proj}}({H}_{\text{struct}})$ \tcp{Structure-conditioned prompting}

    \BlankLine
    \BlankLine
    \tcc{Phase 2: Semantic View (LLM)}
    ${I} = [{P}_{\text{struct}}, \text{Embedding}(\mathcal{T})]$ \tcp{Construct prompted input}

    ${H}_{\text{sem}}, P_{\text{LLM}}
        \leftarrow \text{LLM}_{\theta}({I})$ \tcp{Structure-conditioned forward}

    \BlankLine
    \BlankLine
    \tcc{Phase 3: Dynamic Structure Learning}
    ${S} \leftarrow {H}_{\text{sem}} {W} {H}_{\text{sem}}^{\top}$ \tcp{Compute similarity matrix}

    ${A}_{\text{sem}} \leftarrow \text{Prune}({S}, k)$ \tcp{Semantic graph}

    $\alpha = \sigma(\text{MLP}({H}_{\text{sem}}))$ \tcp{Adaptive gating coefficient}

    $\tilde{{A}} = \alpha \odot {A} + (1 - \alpha) \odot {A}_{\text{sem}}$ \tcp{Semantic-guided graph construction}

    \BlankLine
    \BlankLine
    \tcc{Phase 4: Optimization}
    $\gamma = \sigma\big(\text{MLP}([\mathbb{H}(P_{\text{LLM}}), \mathbb{H}(P_{\text{GNN}}), {H}_{\text{struct}}])\big)$\;

    $Y_{\text{pred}} = \gamma \odot P_{\text{LLM}} + (1 - \gamma) \odot P_{\text{GNN}}$ \tcp{Adaptive gating}

    $\mathcal{L}_{\text{task}}
        \leftarrow \text{CrossEntropy}(Y_{\text{pred}}, \mathcal{Y}_L)$\;

    $\mathcal{L}_{\text{conflict}}$ \tcp{Compute conflict-aware loss using PPR ${\Pi}$}

    $\mathcal{L}_{\text{cons}}$ \tcp{Compute uncertainty-gated consistency loss}

    $\mathcal{L}_{\text{total}}
        \leftarrow \mathcal{L}_{\text{task}}
        + \mathcal{L}_{\text{conflict}}
        + \mathcal{L}_{\text{cons}}$\;

    $\theta_{\text{LLM}}, \theta_{\text{GNN}}
        \leftarrow \text{Backprop}(\mathcal{L}_{\text{total}})$ \tcp{    Update iteratively}
    }

\end{algorithm}

\subsection{Related Work}

\subsubsection{Integration of LLMs and GNNs}
The convergence of Large Language Models and Graph Neural Networks for text-attributed graphs has attracted rapidly growing attention~\cite{ren2024survey,wang2025graph}. Most existing approaches can be broadly classified into three paradigms: LLM-as-Encoder, GNN-as-Prefix, and co-driving. LLM-as-Encoder methods follow a \texttt{Text}$\rightarrow$\texttt{LLM}$\rightarrow$\texttt{GNN} pipeline, where LLMs generate fixed node features consumed by a downstream GNN. TAPE~\cite{he2024harnessing} produces LLM-generated explanations to augment GNN node representations. ZeroG~\cite{li2024zerog} leverages LLM embeddings for zero-shot cross-dataset transfer. OFA~\cite{liu2023one} unifies diverse graph tasks via natural language descriptions, while GAugLLM~\cite{fang2024gaugl} employs LLMs for data augmentation in graph contrastive learning. GNN-as-Prefix methods reverse this direction via a \texttt{GNN}$\rightarrow$\texttt{Projector}$\rightarrow$\texttt{LLM} pipeline, injecting structural encodings into frozen LLMs. GraphGPT~\cite{tang2023graphgpt} employs graph instruction tuning with a GNN encoder. LLaGA~\cite{chen2024llaga} projects GNN representations into the LLM token space, and InstructGLM~\cite{ye2024language} verbalizes graph structure into textual instructions. GraphAdapter~\cite{huang2024can} further explores GNNs as lightweight adapters for frozen LLMs. More recently, co-driving approaches attempt deeper integration: Patton~\cite{jin2023patton} pretrains language models on text-rich networks with graph-aware objectives, FLAG~\cite{yang2025flag} integrates LLMs and GNNs through semantic neighbor sampling for fraud detection, and PromptGFM~\cite{zhu2025llm} combines graph vocabulary learning with LLMs. However, these methods generally follow static pipelines that treat one modality's output as fixed ground truth, preventing mutual error correction. When the upstream model produces hallucinations or noisy encodings, errors propagate irreversibly to the downstream model. In contrast, CO-EVOLVE establishes a cyclic co-evolution loop where the LLM and GNN continuously refine each other via bidirectional feedback.

\subsubsection{Learning on Heterophilous Graphs}
Classical GNNs~\cite{kipf2017semi,velickovic2018graph} implicitly rely on the homophily assumption that connected nodes share similar features and labels. However, this assumption breaks down in heterophilous graphs where neighbors frequently belong to different classes~\cite{gong2026survey,zheng2022graph}. Several architectures have been proposed to address this limitation. H2GCN~\cite{zhu2020beyond} separates ego and neighbor representations and applies intermediate combination strategies to handle non-homophilous patterns. FAGCN~\cite{bo2021beyond} adaptively aggregates low and high-frequency signals via learnable signed attention weights. GPR-GNN~\cite{chien2021adaptive} learns generalized PageRank weights to adapt propagation depths to varying homophily levels, while LINKX~\cite{lim2021large} decouples adjacency and feature information and combines them through separate MLPs for large-scale non-homophilous settings. More recently, Wang et al.~\cite{wang2024understanding} provide theoretical analysis of how different heterophily patterns impact GNN performance. On the contrastive learning front, standard objectives that treat all structural neighbors as positive pairs collapse decision boundaries under heterophily. HLCL~\cite{yang2024hlcl} addresses this by applying low- and high-pass graph filters to construct heterophily-aware contrastive views, and HeterGCL~\cite{wang2024hetergcl} combines adaptive neighbor aggregation with local-to-global contrastive objectives for heterophilic graphs. Despite these advances, existing heterophily methods generally operate within GNN-only frameworks and do not leverage textual semantics to resolve the dissonance between semantic similarity and structural connectivity. Furthermore, they lack explicit mechanisms to mine and penalize hard conflict pairs where textual proximity contradicts topological reality. CO-EVOLVE addresses this gap with a Hard-Structure Conflict-Aware Loss that leverages global PPR diffusion to identify False Semantic Friends and warp the semantic manifold to respect high-order topological boundaries.
\end{document}